\documentclass[draft]{agujournal2018}
\usepackage{apacite}
\usepackage{url} %this package should fix any errors with URLs in refs.
\usepackage{trackchanges}
\usepackage{amsmath}
\usepackage{array}
\usepackage{bm}
\usepackage{booktabs}
\usepackage{color,xcolor}
\allowdisplaybreaks[4]
\usepackage{graphicx}
\usepackage{indentfirst}
\usepackage{multirow}
\usepackage{ragged2e}
\usepackage{supertabular}
\draftfalse

\journalname{JGR: Solid Earth}

\begin{document}

%% ------------------------------------------------------------------------ %%
%  Title
%
% (A title should be specific, informative, and brief. Use
% abbreviations only if they are defined in the abstract. Titles that
% start with general keywords then specific terms are optimized in
% searches)
%
%% ------------------------------------------------------------------------ %%

% Example: \title{This is a test title}

\title{Formulation of a Triaxial Three-Layered Earth Rotation: I. Theory and Rotational Normal Mode Solutions}

%% ------------------------------------------------------------------------ %%
%
%  AUTHORS AND AFFILIATIONS
%
%% ------------------------------------------------------------------------ %%

% Authors are individuals who have significantly contributed to the
% research and preparation of the article. Group authors are allowed, if
% each author in the group is separately identified in an appendix.)

% List authors by first name or initial followed by last name and
% separated by commas. Use \affil{} to number affiliations, and
% \thanks{} for author notes.
% Additional author notes should be indicated with \thanks{} (for
% example, for current addresses).

% Example: \authors{A. B. Author\affil{1}\thanks{Current address, Antartica}, B. C. Author\affil{2,3}, and D. E.
% Author\affil{3,4}\thanks{Also funded by Monsanto.}}

\authors{Zhiliang Guo$^1$, WenBin Shen$^{1,2}$}

% \affiliation{1}{First Affiliation}
% \affiliation{2}{Second Affiliation}
% \affiliation{3}{Third Affiliation}
% \affiliation{4}{Fourth Affiliation}

\affiliation{1}{Department of Geophysics, School of Geodesy and Geomatics/Key Laboratory of Geospace Environment and Geodesy, Wuhan University, Wuhan, 430079, China}

\affiliation{2}{State Key Laboratory of Information Engineering in Surveying, Mapping and Remote Sensing, Wuhan University, Wuhan, 430079, China}
%(repeat as many times as is necessary)

%% Corresponding Author:
% Corresponding author mailing address and e-mail address:

% (include name and email addresses of the corresponding author.  More
% than one corresponding author is allowed in this LaTeX file and for
% publication; but only one corresponding author is allowed in our
% editorial system.)

% Example: \correspondingauthor{First and Last Name}{email@address.edu}

\correspondingauthor{WenBin Shen}{wbshen@sgg.whu.edu.cn}

%% Keypoints, final entry on title page.

%  List up to three key points (at least one is required)
%  Key Points summarize the main points and conclusions of the article
%  Each must be 100 characters or less with no special characters or punctuation

% Example:
% \begin{keypoints}
% \item	List up to three key points (at least one is required)
% \item	Key Points summarize the main points and conclusions of the article
% \item	Each must be 100 characters or less with no special characters or punctuation
% \end{keypoints}

\justifying
\begin{keypoints}
\item A new triaxial three-layered Earth rotation theory was established
\item Four rotational normal modes (CW, FCN, FICN, and ICW) under this new frame were provided
\item FICN can be prograde or retrograde and the ICW period varies significantly due to different core mantle couplings
\end{keypoints}

%% ------------------------------------------------------------------------ %%
%
%  ABSTRACT
%
% A good abstract will begin with a short description of the problem
% being addressed, briefly describe the new data or analyses, then
% briefly states the main conclusion(s) and how they are supported and
% uncertainties.
%% ------------------------------------------------------------------------ %%

%% \begin{abstract} starts the second page

\begin{abstract}
In this study, we formulated a triaxial three-layered anelastic Earth rotation theory considering various core mantle couplings, including the pressure and gravitational couplings acting on the inner core by the outer core and mantle, the viscoelectromagnetic couplings between the outer core and mantle, and between the outer and inner cores. With this formulation, we provided four numerical solutions for the rotational normal modes, including the Chandler Wobble\,(CW), Free Core Nutation\,(FCN), Free Inner Core Nutation\,(FICN), and the Inner Core Wobble\,(ICW). The triaxiality led to increased periods for the CW and ICW of about 0.01 and 0.35 mean solar days (d), respectively. The mantle anelasticity and ocean tide induced dissipations were mainly responsible for the CW, but contributed little to the FCN, while the viscoelectromagnetic coupling induced dissipations were mainly responsible for the FCN, FICN, and ICW. By investigating different types of couplings, we found that pressure coupling played the dominant role in prograde FICN, while viscoelectromagnetic or gravitational couplings either alone, or together gave rise to retrograde FICN. On the other hand, the ICW period varied extensively from 130 d to 21 yr under different core mantle coupling conditions.
\end{abstract}

%% ------------------------------------------------------------------------ %%
%
%  TEXT
%
%% ------------------------------------------------------------------------ %%

%%% Suggested section heads:
% \section{Introduction}
%
% The main text should start with an introduction. Except for short
% manuscripts (such as comments and replies), the text should be divided
% into sections, each with its own heading.

% Headings should be sentence fragments and do not begin with a
% lowercase letter or number. Examples of good headings are:

% \section{Materials and Methods}
% Here is text on Materials and Methods.
%
% \subsection{A descriptive heading about methods}
% More about Methods.
%
% \section{Data} (Or section title might be a descriptive heading about data)
%
% \section{Results} (Or section title might be a descriptive heading about the
% results)
%
% \section{Conclusions}

\section{Introduction}
The study of Earth's rotation is an interdisciplinary pursuit of geodesy, geophysics, and astronomy. Observations from space geodesy and astrometry promote improvements of Earth's rotation theory in geodesy and geophysics. Conventionally, scientists focus on the rotationally symmetric Earth rotation theory, from one-layer to three-layer models \citep{munk1960,lambeck1980,moritz1987,dehant2015} due to the fact that (1) observation accuracies have been historically insufficient, and (2), these models are more easily dealt with. However, with increasing levels of science and technology, observations are achieving successively higher accuracy levels. Hence, a more rigorous triaxial three-layered Earth rotation theory is required. The new theory is expected to better describe the real stratified rotational behaviors of a three-layered Earth, and better reveal the mechanisms of the relevant Earth rotation variation phenomena, such as variations of the Chandler wobble\,(CW) and decadal variations of the length of day\,(LOD). 

Earth's triaxiality is generally neglected in the aforementioned Earth rotational theories. However, the discovery of the Earth's layered lateral heterogeneity from seismic evidence \citep{dziewonski1993,soldati2003,deuss2014} and modern geodetic observations (e.g. \citet{Bursa1992,marchenko2003,groten2004,chen2010}) demonstrate that not only is the whole Earth triaxial, but that the fluid outer core and solid inner core are also triaxial (\citet{chen2015,sun2016}).  \citet{bursa1984} determined the principal moments of inertia for the triaxial Earth using second-order Stokes coefficients and stated that the direction of the largest axis of the best-fitting triaxial Earth's ellipsoid is practically identical with the direction of the axis of the triaxial Earth's smallest principal moments of inertia with a longitude of $-14.90^\circ$. \citet{liu1991} formulated a method for determining the principal inertial axes and the difference between the two equatorial principal moments of inertia of the triaxial Earth. \citet{Bursa1992} computed the principal moments of the triaxial Earth's inertia and their differences from second degree geopotential parameters and astronomical dynamical ellipticity. \citet{marchenko2003} proposed an eigenvalue-eigen vector method to derive a group of principal moments of inertia from second-order gravitational potential coefficients. \citet{groten2004} summarized the principal moments of inertia for the triaxial Earth and other basic parameters related to astronomy, geodesy, and geodynamics. After \citet{chen2010}, \citet{chen2015} further provided the dynamical figure parameters of a triaxial three-layered Earth model, including the principal moments of inertia and dynamical ellipticity for each layer. Hence, to better describe the Earth's rotation behavior it is necessary to consider triaxiality when formulating Earth rotation theories.

Previously, some scientists have studied the triaxial effects based on rigid Earth or elastic Earth models. \citet{seitz2005} studied the atmospheric and oceanic excitations for a triaxial elastic Earth based on their Dynamic Model for Earth Rotation and Gravity\,(DyMEG). \citet{gross2007,gross2015} formulated a triaxial elastic Earth rotation theory and described the observational phenomena of polar motion and length of day variations. \citet{bizouard2013} found that based on a triaxial elastic Earth model, the pole tide and the triaxiality will both lead to asymmetric polar motion, while the effect of triaxiality will partly cancel the effect of the pole tide. Though \citet{hinderer1982} formulated a triaxial two-layered Earth rotation theory, they studied the perturbations of the CW eigenfrequencies and the Nearly Diurnal Free Wobble\,(NDFW) under the rotationally symmetric condition. \citet{hoolst2002} successfully considered the triaxial and second-order effects of geometric and dynamical flattenings on the normal modes CW and Free Core Nutation\,(FCN) of a two-layered Earth model and Mars model under the framework of \citet{dehant1993}. 

 In the last ten years, \citet{shen2007} and \citet{shen2008} investigated the rotational behavior of a triaxial rigid Earth and estimated the principal moments of inertia for the whole Earth. \citet{chen2010} formulated a triaxial two-layered Earth rotational theory and provided the normal mode solutions that match the observations very well after considering the mantle anelasticity and ocean tide dissipations, and applied the theory to polar motion excitation while considering a frequency dependent response. \citet{chen2013a} and \citet{chen2013b} further investigated the frequency dependent response of polar motion excitation by modeling the mantle anelasticity, quasi-fluid rheology, ocean tides, and core mantle coupling, and \citet{chen2015} estimated the triaxial dynamic figure parameters of the triaxial three-layered Earth based on new gravity field models EGM2008, EIGEN-6C, and EIGEN-6C2. \citet{sun2016} extended the theory of \cite{chen2010} by considering the complex compliances and inverted the core triaxiality parameters by using the difference between the empirical and theoretical models for prograde diurnal polar motion. Under the triaxial two-layered Earth rotation frame, \citet{yang2016} further considered the electromagnetic coupling between the mantle and fluid core and investigated the effects of the compliances on the rotational CW and FCN normal modes.
 
 Before formulating a triaxial three-layered Earth rotation theory and investigating the effects of triaxiality on the rotational normal modes, we first reviewed the current rotationally symmetric Earth rotation theories based on the angular momentum balance (AMB) approach that was used in our formulation. Due to the pioneer work of \citet{hough1895} and \citet{poincare1910} for the rigid mantle and incompressible fluid core, \citet{SOS1977} built a rotationally symmetric two-layered Earth rotation theory by considering the elastic mantle and incompressible fluid core, and \citet{SOS1980} extended the model by considering a compressible fluid core instead. Since the moments of inertia of the inner core are only about 7/10000 of the whole Earth, the inner core is often neglected in conventional Earth rotation theory. The two-layered Earth rotational theory of \citet{hinderer1982} is different from the work of \citet{SOS1977} and \citet{SOS1980}, and the former was applied by \citet{cui2012} to study the FCN and core mantle coupling effects. \citet{jochmann2009} formulated a two-layered Earth rotational model with inertia coupling and provided the rotational normal mode analytical solutions for CW and FCN, and investigated the effect of pole tide on the CW. \citet{szeto1984a} and \citet{smylie1984} deployed the Euler kinematic and dynamic equations to describe the motion of the inner core, and found a rapid precession and a slow precession in the mantle frame that were later confirmed as the Free Inner Core Nutation\,(FICN) and Inner Core Wobble\,(ICW) \citep{Xu1998}. \citet{mathews1991a} formulated a rotationally symmetric three-layered Earth rotation theory with pressure and gravitational couplings considered\,(simply MBHS1991 theory), and stated that two more normal modes will be induced because of the inner core, namely the FICN and ICW. Further, \citet{mathews2002} considered the effects of electromagnetic couplings, mantle anelasticity, and ocean tide dissipations, finding that the FICN period will be prolonged to 1025 mean solar days\,(d). In the frame of the AMB approach, \citet{dehant1993} also independently formulated a rotationally symmetric three-layered Earth rotation theory, and provided four rotational normal modes. \citet{legros1993} accounted for the resonance effects induced by the tidal potential, atmospheric loading in the tides, and nutation based on the work of \citet{dehant1993}. Under the framework of MBHS1991 theory, \citet{dumberry2002} investigated the effect of the inner core tilt on decadal polar motion excitation that is frequently referred to as the Markowitz wobble, and found that a torque of $10^{20}$ N m with a 0.07 degree inner core tilt is needed that critically depends on the viscosity of the inner core. Later, \citet{dumberry2008a} further explained the Markowitz wobble based on a time-dependent axial misalignment model between the density structures of the inner core and mantle. \citet{dumberry2008b} and \citet{dumberry2009} successfully extended the MBHS1991 theory to consider the inner core tilt induced elastic deformations, while \citet{dumberry2016} extended the MBHS1991 theory to formulate a new rotation model to study the forced precession of the inner core of the Moon. 

Generally, there are two kinds of normal modes in solid Earth system theories \citep{chao2017}, seismic normal modes (or free oscillation modes) mainly excited episodically by earthquakes \citep{dahlen1968,dahlen1969}, and rotational normal modes continually excited by mass redistributions, relative motions, and external torques \citep{smith1977,wahr1981a,wahr1981c,mathews1991a,mathews1991b,dehant1993,hoolst2002,mathews2002,rogister2009,chen2010,rochester2014,crossley2014}. The normal modes reflect the structure, components, and physics of the Earth. For example, various free oscillation modes (combined with other observations) were used to establish the preliminary reference Earth model (PREM) \citep{dziewonski1981} that has broad applications in the geosciences. In another aspect, the rotational normal modes can be used to reveal the rotational behavior and physics of the stratified Earth, such as excitation mechanisms and core mantle coupling properties \citep{cui2012}.
 
Hence, scientists pay great attention to various rotational normal modes for the Earth, including the CW, FCN, FICN, and ICW. Various observations of the rotational CW and FCN normal modes have been obtained by analyzing different kinds of data sets, including for instance, the Earth orientation parameter (EOP) data, very long baseline interferometry (VLBI) data, superconducting gravity data, and others. \citet{furuya1996} estimated the period and quality factor Q of the CW with polar motion data and Atmospheric Angular Momentum\,(AAM) data by investigating the Inverted Barometer\,(IB) effect on excitation functions, while \citet{vondrak2017} determined the period and Q of the CW based on numerical integration of the broad-band Liouville equations by considering the effect of geomagnetic jerk. \citet{nastula2015} obtained the period and Q of the CW by minimizing the difference between the modeled and observed polar motion excitation functions that are derived from polar motion data and second degree gravitational potential coefficients based on observations of satellite laser range (SLR) (e.g. \citet{cheng2011}) and Gravity Recovery and Climate Experiment (GRACE) (e.g., \citet{bettadpur2012}). Due to time-varying excitation and damping, the FCN is not included in precession-nutation model and remains in the celestial pole offsets\,(CPO) data observed by the VLBI technique. Hence, the period and Q of the FCN can be analyzed with these data by using different methods \citep{lambert2007,krasna2013,chao2015,zhou2016}. \citet{chao2015} formulated the dynamics of the FCN and estimated its eigenperiod using VLBI data. Observed from the Terrestrial Reference System, the FCN will become the NDFW, and due to strong resonance with the solid Earth tides in the diurnal band, the NDFW can be observed using superconducting gravity data \citep{ducarme2007,rosat2009b}. \citet{rosat2009a} and \citet{rosat2017} successfully obtained the FCN parameters using both the VLBI data and superconducting gravity data. Recently, \citet{rosat2016} and \citet{rosat2017} searched for a possible FICN resonance effect with a period of 1300 sidereal days using superconducting gravimeter data and the CPO series. However, until now, there are no reliable observations of the periods and Qs of the FICN and ICW, and this attracts great interests from scientists.

In theory, to solve the rotational normal modes of Earth's rotation, generally three kinds of methods are used: the AMB approach \citep{SOS1977,SOS1980,SW1981,hinderer1982,mathews1991a,dehant1993,mathews2002}, displacement field\,(DF) approach \citep{smith1974,wahr1981a,deVries1991,rogister2009,rochester2014,dehant2015}, and the Hamiltonian analysis\,(HA) approach \citep{getino-jose1997,getino-jose2000,ferrandiz2015}. Based on the AMB approach, the rigid Earth model provides one rotational normal mode, namely the CW \citep{moritz1987}. The two-layered Earth model with a rigid or elastic mantle and an incompressible or compressible fluid core will generate two normal modes, namely the CW and FCN, with different periods \citep{SOS1977,SOS1980}. The biaxial three-layered Earth model with an elastic solid mantle, compressible fluid outer core, and solid inner core will give rise to four normal modes, namely the CW, FCN, FICN, and ICW \citep{mathews1991a,dehant1993,legros1993}. Based on the DF approach, except the rotational normal mode CW, as summarized by \citet{mathews1991a}, the FCN, FICN and ICW normal modes will appear if the fluid core and solid inner core are considered. For example, by considering the solid inner core and its non-hydrostatic structure, \citet{deVries1991} independently found the new normal mode FICN. \citet{rogister2009} successfully investigated the influences of the liquid outer core on the CW, FCN, and FICN rotational normal modes. \citet{crossley2014} studied the CW, FCN, FICN and ICW rotational normal modes based on their wobble-nutation theory \citep{rochester2014}. Based on the HA approach, for a two-layered solid mantle and fluid core Earth model, there will be two rotational normal modes, namely, the CW and FCN \citep{gonzalez1997,getino2000}, and for a three-layered solid mantle, fluid outer core, and solid inner core Earth model, there will be four rotational normal modes, the CW, FCN, FICN, and ICW \citep{escapa2001}, consistent with the AMB approach. 

Here, aligned with previous studies \citep{mathews1991a,chen2010,sun2016,yang2016}, we formulated a triaxial three-layered Earth rotation theory. The Earth is composed of an anelastic mantle (including the crust, hereafter, the same meaning), a fluid outer core, and a solid inner core. In section 2, we first generalized the MBHS1991 theory by extending the rotational symmetric moments of the inertial tensor into the triaxial tensors. Then, the external tesseral tidal, pressure, and gravitational coupling torques were extended into the triaxial theory to match the triaxial moments of inertia tensor. In addition, the viscoelectromagnetic coupling torques near the core-mantle boundary\,(CMB) and inner core boundary\,(ICB) were considered and extended into the triaxial theory. A new triaxial three-layered Earth rotation theory was then established. In section 3, we used two approaches to solve the normal modes of the triaxial three-layered Earth rotation equations. In section 4, we describe the needed input data, including the dynamical figure parameters of the triaxial three-layered Earth, compliances partly from \citet{yang2016} and partly from \citet{mathews2002}, and the core mantle coupling parameters. In section 5, to validate the triaxial three-layered Earth rotation theory, we calculated the corresponding four rotational normal modes, and compared the new results with those obtained by conventional theories. Additionally, the dissipative properties of the four normal Earth rotation modes are discussed. Since there are lots of observations of the CW and FCN periods, here we used the means of the CW and FCN period observations to adjust the values of the compliances to best match the observations. In section 6, we studied the rotational normal modes extensively by investigating the rotational normal modes under different core mantle coupling conditions with some simplified cases, because different core mantle couplings, like viscoelectromagnetic, gravitational, pressure couplings will have significant and distinct effects on the rotational normal modes, i.e. the CW, FCN, FICN and ICW. In section 7 we summarized the study and discuss relevant problems. 
  
%------------------------------------------------

\section{Theory}
\label{S2}
In this study we adopted the basic assumptions deployed by \citet{mathews1991a} and \citet{mathews2002}. The standard Earth, composed of the elastic mantle, fluid outer core, and the solid inner core are defined as rotating with a mean angular velocity $\bm{\Omega_{0}}$ about a mean rotational vector $\mathbf{I_{3}}$ along the $\mathbf{z}$ axis of a geocentric quasi-inertial reference system called the geocentric celestial reference system\,(GCRS). As discussed in \citet{mathews1991a}, the displacement field of the real Earth with respect to the standard Earth can be divided into a rigid rotation component and a deformation field. We applied the rigid rotation to the solid inner core of the standard Earth to define the instantaneous figure axis $\mathbf{i'_3}$, and also applied the rigid rotation to the mantle and fluid outer core of the standard Earth to define the mantle fixed $\mathbf{i}$-system $(\mathbf{i_{1}},\mathbf{i_{2}},\mathbf{i_{3}})$. These two coordinate systems were designed to separate the rigid rotation induced inertial tensor components and the deformations induced by varying the inertial tensor components, where the rigid rotation induced components constitute two independent quasi-principal axis systems, and the deformations caused by tides and non-uniform rotation are considered later as deviations. The tilt of the inner core is thus $\mathbf{n_{s}}=\mathbf{i'_{3}}-\mathbf{i_{3}}$. The $\mathbf{i}$-system is close to, but not the Tisserand mean axial system (TMAS) \citep{munk1960}, as stated by \citet{mathews1991a}. Thus, in an inertial system, the angular velocities of the mantle, fluid outer core, and solid inner core can be expressed as \citep{mathews1991a}: 
 \begin{equation}
 \begin{gathered}
  \label{1}
  \bm{\Omega}=\bm{\Omega_{0}}+\bm{\omega}=\Omega_{0}(\mathbf{i_{3}}+\bm{m})\\
  \bm{\Omega_{f}}=\bm{\Omega}+\bm{\omega_{f}}=\Omega_{0}(\mathbf{i_{3}}+\bm{m}+\bm{m_{f}})\\
  \bm{\Omega_{s}}=\bm{\Omega}+\bm{\omega_{s}}=\Omega_{0}(\mathbf{i_{3}}+\bm{m}+\bm{m}_{s}).
 \end{gathered}
 \end{equation}
 
 In the MBHS1991 theory, only terms with a magnitude of $O(m)$ or $O(m\epsilon)$ are considered, and the 2nd order terms with magnitudes of $O(m^2)$ and $O(m\epsilon^2)$ or higher order terms are neglected. The pressure and gravitational coupling torques acting on the fluid outer core are neglected in the fluid outer core rotational equation (see equation (\ref{3})) due to the higher order effect of $O(m\epsilon^2)$. This is referred to as \emph{SOS approximation}, that was first proposed by \citet{SOS1980} and later used by \citet{mathews1991a}. Similarly, in our study, we also considered only the first order terms of $O(m)$ or $O(m\epsilon)$, and neglected the higher order terms for simplicity. In addition, \citet{mathews2005} extended the electromagnetic coupling models in \citet{mathews2002} to include viscous couplings, thus the new viscoelectromagnetic coupling torques were generalized and applied to the triaxial case. We note that considering higher order terms in triaxial three-layered rotation theory is much more complicated. For example, the second order dynamical and geometric flattening effects of a triaxial three-layered Earth with a magnitude of $O(m\epsilon^2)$, or the second order rotational parameter effects of a triaxial three-layered Earth with a magnitude of $O(m^2)$ were not considered here. 

Based on the above assumptions, the coupled rotational equation for a triaxial three-layered Earth can be formulated in the $\mathbf{i}\,(\mathbf{i_{1}},\mathbf{i_{2}},\mathbf{i_{3}})$ principal axis system \citep{mathews1991a,mathews2002,dumberry2008a,koot2008,dumberry2009}: 
\begin{eqnarray}
 \label{2}
 \frac{\mathrm{d}\bm{H}}{\mathrm{d}t}+\bm{\Omega}\times\bm{H}=\bm{\Gamma}\\
 \label{3}
 \frac{\mathrm{d}\bm{H}_f}{\mathrm{d}t}-\bm\omega_f\times\bm{H}_f=\bm{\Gamma}_{CMB}-\bm{\Gamma}_{ICB}\\
 \label{4}
 \frac{\mathrm{d}\bm{H}_s}{\mathrm{d}t}+\bm{\Omega}\times\bm{H}_s=\bm{\Gamma}_{s}+\bm{\Gamma}_{ICB}
\end{eqnarray}
where $\bm{H}$ and $\bm{\Omega}$ are the angular momentum and angular velocity for the whole Earth, respectively. $\bm{\Gamma}$ is the external tesseral tidal torque, $\bm{H}_f$ is the angular momentum of the fluid outer core, $\bm{\omega}_f$ is the angular velocity of the outer core with respect to the mantle, $\bm{\Gamma}_{CMB}$ is the viscoelectromagnetic coupling torque acting on the fluid outer core by the mantle, $\bm{\Gamma}_{ICB}$ is the viscoelectromagnetic coupling torque acting on the solid inner core by the fluid outer core, $\bm{H}_s$ is the angular momentum of the solid inner core, and $\bm{\Gamma}_s$ is the pressure and gravitational coupling torque acting on the solid inner core by the fluid outer core and mantle. 

Following \citet{mathews1991a}, the external tesseral tidal torque acting on the whole Earth, $\bm{\Gamma}$, can be extended into the triaxial case, expressed as
\begin{equation}
\label{5}
\bm{\Gamma}=\Omega^2_0 \left( \begin{array}{c}
\vspace{3mm}
 (C-B)\phi_2 \\
 \vspace{3mm}
 -(C-A)\phi_1 \\
 0
 \end{array}\right)
\end{equation}
where $A$, $B$, $C$ are the moments of inertia for the whole Earth, and $\phi=\phi_1+i \phi_2$ is the dimensionless tesseral tidal potential expressed as
\begin{eqnarray}
 \label{6}
 \phi_1 =\frac{3GM_c}{\Omega^2_0 d^5}d_xd_z \\
 \label{7}
 \phi_2=\frac{3GM_c}{\Omega^2_0 d^5}d_yd_z
 \end{eqnarray}
 where $M_c$ is the mass of the perturbing celestial body, namely the moon, the Sun, or other planets in our solar system, and $(d_x,d_y,d_z)$ are the components of the position vector ${\bf d}$ for that celestial body. 

According to \citet{mathews2002} and \citet{mathews2005}, the viscoelectromagnetic coupling torque acting on the fluid outer core by the mantle, $\bm{\Gamma}_{CMB}$, can be expressed in the triaxial form as
\begin{equation}
\label{8}
\bm{\Gamma}_{CMB}=K_{CMB} \Omega^2_0 \left( \begin{array}{c}
\vspace{3mm}
B_f m^f_2 \\
\vspace{3mm}
-A_f m^f_1 \\
0
\end{array}\right)
\end{equation}
where $K_{CMB}$ is the dimensionless viscoelectromagnetic coupling parameter between the mantle and fluid outer core, $A_f$ and $B_f$ are the equatorial moments of inertia for the fluid outer core, $m^f_1$ and $m^f_2$ are the equatorial components for the rotational parameters of the fluid outer core and the viscoelectromagnetic coupling torque acting on the solid inner core by the fluid outer core. $\bm{\Gamma}_{ICB}$ can be expressed in the triaxial form as 
\begin{equation}
\label{9}
\bm{\Gamma}_{ICB}=K_{ICB} \Omega^2_0 \left( \begin{array}{c}
\vspace{3mm}
B_s (m^s_2-m^f_2) \\
\vspace{3mm}
-A_s(m^s_1-m^f_1) \\
0
\end{array}\right)
\end{equation}
where $K_{ICB}$ is the dimensionless viscoelectromagnetic coupling parameter between the fluid outer core and solid inner core, $A_s$ and $B_s$ are the equatorial moments of inertia for the solid inner core, and $m^s_1$ and $m^s_2$ are the equatorial components of the rotational parameters for the solid inner core. 

The pressure and gravitational coupling torque acting on the solid inner core by the fluid outer core and mantle, $\bm{\Gamma}_s$, can also be extended into the triaxial cases, expressed as
\begin{equation}
\label{10}
\bm{\Gamma}_s=\Omega^2_0 \left( \begin{array}{c} 
\vspace{3mm}
(C_s-B_s)\big[\alpha_1 (m_2+m^f_2)-\alpha_2 n^s_2 + \alpha_3 \phi_2 \big] - c^s_{23} \\
\vspace{3mm}
-(C_s-A_s)\big[\alpha_1 (m_1+m^f_1)-\alpha_2 n^s_1 + \alpha_3 \phi_1 \big] + c^s_{13} \\
0
\end{array}\right)
\end{equation}
compared to the rotationally symmetric cases given by \citet{mathews1991a}. In equation (\ref{8}), $A_s$, $B_s$, and $C_s$ are the moments of inertia for the solid inner core, $\alpha_1$, $\alpha_2$, and $\alpha_3$ are the pressure and gravitational coupling parameters, $\mathbf{n_{s}}=\mathbf{i'_{3}}-\mathbf{i_{3}}=(n^s_1,n^s_2,n^s_3)$ is the tilt of the solid inner core and $c^s_{13}$ and $c^s_{23}$ are inertia products induced by elastic deformations of the solid inner core due to tides and non-uniform rotation. The pressure and gravitational coupling parameters $\alpha_1$, $\alpha_2$, and $\alpha_3$ can be represented as \citep{mathews1991a}:
\begin{equation}
\begin{gathered}
 \alpha_1=1-\alpha_3=\frac{C'_s-(A'_s+B'_s)/2}{C_s-(A_s+B_s)/2} \\
 \label{11}
 \alpha_2=\alpha_1-\alpha_3 \alpha_g \\
 \alpha_g=\frac{8\pi G}{5 \Omega^2_0}\left(\int^a_{a_s}\rho_0(a')\frac{\mathrm{d}\epsilon}{\mathrm{d}a'}da'+\rho_f \epsilon_s \right)
\end{gathered}
\end{equation}
where $A'_s$, $B'_s$, and $C'_s$ are the corresponding moments of inertia for the inner core that are replaced by an inner core with density $\rho_f$ that is also the density of the fluid outer core at the ICB, $\alpha_g$ represents the strength of the gravitational coupling, $\epsilon_s$ is the geometric flattening of the solid inner core, $\epsilon$ is the geometric flattening of the fluid outer core and the mantle varied from the ICB to the surface of the solid Earth, and $a_s$ and $a$ are the radii of the solid inner core and solid Earth, respectively. 

Since our present study focuses only on polar motion and nutation, we did not consider the z-components of the core mantle coupling torques (including for instance, the z-components of the electromagnetic coupling torques near the CMB and ICB and the pressure and gravitational coupling torque). In addition, the z-component of the tidal torque is the second order sectorial tidal potential, hence it was neglected in this study. 

The angular momenta of the solid inner core, the fluid outer core, and whole  Earth can be written as \citep{mathews1991a}:
 \begin{eqnarray}
  \label{12}
   &\mathbf{H}_{s}=[\mathbf{C}_{s}]\cdot \bm{\Omega_{s}}\\
   \label{13}
   &\mathbf{H}_{f}=[\mathbf{C}_{f}]\cdot \bm{\Omega_{f}}\\
   \label{14}
   &\mathbf{H}=[\mathbf{C}]\cdot \bm{\Omega} + [\mathbf{C}_{f}]\cdot (\bm{\Omega_{f}}-\bm{\Omega})+[\mathbf{C}_{s}]\cdot (\bm{\Omega_{s}}-\bm{\Omega})+\mathbf{H}^{(R)},
  \end{eqnarray}
where the TMAS is used for the fluid outer core and solid inner core, and the coordinate system fixed with the mantle is not, but close to the TMAS, as discussed in section \ref{S2}. This is the reason why the relative angular momentum $\mathbf{H}^{(R)}$ variable exists in the equation for the whole  Earth and can be neglected in the following derivations in our first-order approximation. Following \citet{mathews1991a}, in the triaxial case, the moments of inertia tensor can be generalized as follows:
  \begin{eqnarray}
   \label{15} 
   [\mathbf{C}_{s}]= A_s\mathbf{i_1i_1}+B_s\mathbf{i_2i_2}+C_s\mathbf{i_3i_3}+\big(C_s- \frac{A_s+B_s}{2}\big)(\mathbf{i'_3i'_3}-\mathbf{i_3i_3})+\sum_{ij}c^s_{ij}\mathbf{i}_{i}\mathbf{i}_{j}\\
   \label{16}
   [\mathbf{C}_{f}]=A_f\mathbf{i_1i_1}+B_f\mathbf{i_2i_2}+C_f\mathbf{i_3i_3}+\big(C'-\frac{A'_s+B'_s}{2} \big)(\mathbf{i_3i_3}-\mathbf{i'_3i'_3})+\sum_{ij}c^f_{ij}\mathbf{i}_{i}\mathbf{i}_{j}\\
    \label{17}
   [\mathbf{C}]= A\mathbf{i_1i_1}+B\mathbf{i_2i_2}+C\mathbf{i_3i_3} 
    +\Big[\big(C_s+\frac{A_s+B_s}{2} \big)-\big(C'-\frac{A'_s+B'_s}{2} \big) \Big] (\mathbf{i'_3i'_3}-\mathbf{i_3i_3})\nonumber\\
    +\sum_{ij}c_{ij}\mathbf{i}_{i}\mathbf{i}_{j},
  \end{eqnarray}
where $c^s_{ij}$, $c^f_{ij}$, and $c_{ij}$ are the moments and products of inertia caused by elastic deformation of the solid inner core, fluid outer core, and the whole Earth due to tidal and centrifugal potential, respectively. 

Next, we derived the rotational equations for the whole Earth, fluid outer core, and solid inner core. In the following derivation, we treated $\bm{m}$, $\bm{m}_f$, $\bm{m}_s$, $\mathbf{n_s}$, $c^s_{ij}$, $c^f_{ij}$, and $c_{ij}$ as first order small quantities, and the higher orders were neglected.

Substituting the moments of inertia tensors $[\mathbf{C}_s]$, $[\mathbf{C}_f]$, and $[\mathbf{C}]$ as expressed by equations (\ref{15}), (\ref{16}), and (\ref{17}) into equation (\ref{14}), the angular momentum of the whole Earth $\bm{H}$ can be obtained. Combined with the external tidal torque $\bm{\Gamma}$ as expressed by equation (\ref{5}), the rotational equation for the whole Earth, expressed by (\ref{2}), can be re-formulated as:
  \begin{eqnarray}
  \nonumber
  A\frac{\mathrm{d}m_1}{\mathrm{d}t}+A_f\frac{\mathrm{d}m^f_1}{\mathrm{d}t}+A_s\frac{\mathrm{d}m^s_1}{\mathrm{d}t}+\frac{A_s+B_s}{2}\alpha_3 e_s \frac{\mathrm{d}n^s_1}{\mathrm{d}t}+\frac{\mathrm{d}c_{13}}{\mathrm{d}t}\\ 
  \label{18}
  +\Omega_0 \big((C-B)m_2-(B_f m^f_2+B_s m^s_2+\frac{A_s+B_s}{2}\alpha_3 e_s n^s_2+c_{23})\big)=\Omega_0 (C-B)\phi_2\\
  \nonumber
  B\frac{\mathrm{d}m_2}{\mathrm{d}t}+B_f\frac{\mathrm{d}m^f_2}{\mathrm{d}t}+B_s\frac{\mathrm{d}m^s_2}{\mathrm{d}t}+\frac{A_s+B_s}{2}\alpha_3 e_s \frac{\mathrm{d}n^s_2}{\mathrm{d}t}+\frac{\mathrm{d}c_{23}}{\mathrm{d}t}\\
  \label{19}
  +\Omega_0 \big(-(C-A)m_1+(A_f m^f_1+A_s m^s_1+\frac{A_s+B_s}{2}\alpha_3 e_s n^s_1+c_{13}) \big)=-\Omega_0 (C-A)\phi_1\\
  \label{20}
  C\frac{\mathrm{d}m_3}{\mathrm{d}t}+C_f\frac{\mathrm{d}m^f_3}{\mathrm{d}t}+C_s\frac{\mathrm{d}m^s_3}{\mathrm{d}t}+\frac{A_s+B_s}{2}\alpha_3 e_s \frac{\mathrm{d}n^s_3}{\mathrm{d}t}+\frac{\mathrm{d}c_{33}}{\mathrm{d}t}=0.
  \end{eqnarray}
 Substituting the moments of inertia tensor $[\mathbf{C}_f]$ as expressed by equation (\ref{16}) into equation (\ref{13}), and further substituting equations (\ref{13}), (\ref{8}), and (\ref{9}) into equation (\ref{3}), the rotational equation of the fluid outer core can be obtained as:
  \begin{eqnarray}
    \nonumber
    A_f\frac{\mathrm{d}}{\mathrm{d}t}(m_1+m^f_1)-\frac{A_s+B_s}{2}\alpha_1 e_s \frac{\mathrm{d}n^s_1}{\mathrm{d}t}+\frac{\mathrm{d}c^f_{13}}{\mathrm{d}t}\\ \label{21}
    -\Omega_0(C_f+K_{CMB}B_f+K_{ICB}B_s)m^f_2+\Omega_0 K_{ICB}B_s m^s_2=0 \\
    \nonumber
    B_f\frac{\mathrm{d}}{\mathrm{d}t}(m_2+m^f_2)-\frac{A_s+B_s}{2}\alpha_1 e_s \frac{\mathrm{d}n^s_2}{\mathrm{d}t}+\frac{\mathrm{d}c^f_{23}}{\mathrm{d}t}\\ \label{22}
    +\Omega_0(C_f+K_{CMB}A_f+K_{ICB}A_s)m^f_1-\Omega_0 K_{ICB}A_s m^s_1=0 \\
    \label{23}
    C_f\frac{\mathrm{d}}{\mathrm{d}}(m_3+m^f_3)-\frac{A_s+B_s}{2}\alpha_1 e_s \frac{\mathrm{d}n^s_3}{\mathrm{d}t}+\frac{\mathrm{d}c^f_{33}}{\mathrm{d}t}=0.
  \end{eqnarray}
 Substituting the moments of inertia tensor $[\mathbf{C}_s]$ into equation\,(\ref{12}), and further by substituting equations (\ref{12}), (\ref{9}), and (\ref{10}) into equation (\ref{4}), the rotational equation of the solid inner core can be obtained as:
  \begin{eqnarray}
    \nonumber
    A_s\frac{\mathrm{d}}{\mathrm{d}t}(m_1+m^s_1)+\frac{A_s+B_s}{2}e_s\frac{\mathrm{d}n^s_1}{\mathrm{d}t}+\frac{\mathrm{d}c^s_{13}}{\mathrm{d}t}+\Omega_0 \big((C_s-B_s)m_2+B_s K_{ICB} m^f_2-B_s(1+K_{ICB})m^s_2\\
    \label{24}
    -\frac{A_s+B_s}{2}e_s n^s_2-c^s_{23}\big)=\Omega_0(C_s-B_s)[\alpha_1(m_2+m^f_2)-\alpha_2 n^s_2+\alpha_3\phi_2]-\Omega_0 c^s_{23} \\
    \nonumber
    B_s\frac{\mathrm{d}}{\mathrm{d}t}(m_2+m^s_2)+\frac{A_s+B_s}{2}e_s\frac{\mathrm{d}n^s_2}{\mathrm{d}t}+\frac{\mathrm{d}c^s_{23}}{\mathrm{d}t}+\Omega_0 \big(-(C_s-A_s)m_1-A_s K_{ICB} m^f_1 +A_s(1+K_{ICB})m^s_1\\
    \label{25}
    +\frac{A_s+B_s}{2}e_s n^s_1+c^s_{13} \big)=-\Omega_0(C_s-A_s)[\alpha_1(m_1+m^f_1)-\alpha_2 n^s_1+\alpha_3\phi_1]+\Omega_0 c^s_{13}\\
    \label{26}
    C_s \frac{\mathrm{d}}{\mathrm{d}t}(m_3+m^s_3)+\frac{A_s+B_s}{2}e_s\frac{\mathrm{d}n^s_3}{\mathrm{d}t}+\frac{\mathrm{d}c^s_{33}}{\mathrm{d}t}=0.
  \end{eqnarray}
  In addition, the tilt equation of the solid inner core in the first order can be written as \citep{mathews1991a,guo2002}:
  \begin{eqnarray}
    \label{27}
    \frac{\mathrm{d}n^s_1}{\mathrm{d}t}-\Omega_0 m^s_2=0\\
    \label{28}
    \frac{\mathrm{d}n^s_2}{\mathrm{d}t}+\Omega_0 m^s_1=0\\
    \label{29}
    \frac{\mathrm{d}n^s_3}{\mathrm{d}t}=0
  \end{eqnarray}

  The rotational equations for the whole Earth, fluid outer core, and solid inner core are expressed by equations\,(\ref{18})-(\ref{26}), along with the tilt equations of the solid inner core that are expressed by equations (\ref{27})-(\ref{29}), are the foundation of the triaxial three-layered Earth rotational theory. In the following sections, the rotational normal modes for the triaxial three-layered Earth rotation equations that are related with the wobbles and nutations will be investigated, . Here, we deployed the first two equations in each group of the rotational equations to focus on nutation and polar motion in this study, while the LOD problem will be investigated in a separate study. 
  
  Notice that the inertial products $c_{13}$, $c_{23}$, $c^f_{13}$, $c^f_{23}$, $c^s_{13}$, and $c^s_{23}$ are caused by elastic deformations due to tides and non-uniform rotation, and can be modeled in the same way as \citet{mathews1991a} and then extended into the triaxial case as:
  \begin{eqnarray}
   \label{30}
   c_{13}=A[\kappa(m_1-\phi_1)+\xi m^f_1+\varsigma m^s_1] \\
   \label{31}
   c_{23}=B[\kappa(m_2-\phi_2)+\xi m^f_2+\varsigma m^s_2]\\
   \label{32}
   c^f_{13}=A_f[\gamma(m_1-\phi_1)+\beta m^f_1+\delta m^s_1]\\
   \label{33}
   c^f_{23}=B_f[\gamma(m_2-\phi_2)+\beta m^f_2+\delta m^s_2]\\
   \label{34}
   c^s_{13}=A_s[\theta(m_1-\phi_1)+\chi m^f_1+\nu m^s_1]\\
   \label{35}
   c^s_{23}=B_s[\theta(m_2-\phi_2)+\chi m^f_2+\nu m^s_2]
  \end{eqnarray}
where the compliances $\kappa$,\,$\xi$,\, and $\varsigma$ contribute to the deformational products of inertia $c_{13}$ and $c_{23}$ for the whole Earth, compliances $\gamma$,\,$\beta$,\,and $\delta$ contribute to the deformational products of inertia $c^f_{13}$ and $c^f_{23}$ for the fluid outer core, compliances $\theta$,\,$\chi$,\,and $\nu$ contribute to the deformational products of inertia $c^s_{13}$ and $c^s_{23}$ for the solid inner core that can be calculated based on the elastic displacement fields in a similar way as \citet{SOS1980} and \citet{mathews1991a}. The actual values of these compliances are provided in section \ref{4}. 

Combining equations (\ref{18}), (\ref{19}), (\ref{30}), and (\ref{31}), the rotational equation for the whole Earth can be expressed as:
  \begin{eqnarray}
    \nonumber
    (1+\kappa)A\frac{\mathrm{d}m_1}{\mathrm{d}t}+(A_f+A\xi)\frac{\mathrm{d}m^f_1}{\mathrm{d}t}+(A_s+A\varsigma)\frac{\mathrm{d}m^s_1}{\mathrm{d}t}+\frac{A_s+B_s}{2}\alpha_3 e_s \frac{\mathrm{d}n^s_1}{\mathrm{d}t}\\
    \nonumber
    +(C-(1+\kappa)B)\Omega_0 m_2-(B_f+B\xi)\Omega_0 m^f_2-(B_s+B\varsigma)\Omega_0 m^s_2-\frac{A_s+B_s}{2}\alpha_3 e_s \Omega_0 n^s_2 \\
    \label{36}
    =A\kappa\frac{\mathrm{d}\phi_1}{\mathrm{d}t}+(C-(1+\kappa)B)\Omega_0 \phi_2 \hspace{2mm},\\
    \nonumber
    (1+\kappa)B\frac{\mathrm{d}m_2}{\mathrm{d}t}+(B_f+B\xi)\frac{\mathrm{d}m^f_2}{\mathrm{d}t}+(B_s+B\varsigma)\frac{\mathrm{d}m^s_2}{\mathrm{d}t}+\frac{A_s+B_s}{2}\alpha_3 e_s \frac{\mathrm{d}n^s_2}{\mathrm{d}t}\\
    \nonumber
    -(C-(1+\kappa)A)\Omega_0 m_1+(A_f+A\xi)\Omega_0 m^f_1+(A_s+A\varsigma)\Omega_0 m^s_1+\frac{A_s+B_s}{2}\alpha_3 e_s \Omega_0 n^s_1 \\
    \label{37}
    =B\kappa\frac{\mathrm{d}\phi_2}{\mathrm{d}t}-(C-(1+\kappa)A)\Omega_0 \phi_1 \hspace{2mm};
  \end{eqnarray}
  combining equations (\ref{21}), (\ref{22}), (\ref{32}) and (\ref{33}), the rotational equation of the fluid core can be expressed as:
  \begin{eqnarray}
  \nonumber
  A_f(1+\gamma)\frac{\mathrm{d}m_1}{\mathrm{d}t}+A_f(1+\beta)\frac{\mathrm{d}m^f_1}{\mathrm{d}t}+A_f\delta\frac{\mathrm{d}m^s_1}{\mathrm{d}t}-\frac{A_s+B_s}{2}\alpha_1 e_s \frac{\mathrm{d}n^s_1}{\mathrm{d}t} \\
  \label{38}
  -\Omega_0(C_f+K_{CMB}B_f+K_{ICB}B_s)m^f_2+\Omega_0 K_{ICB}B_s m^s_2=A_f \gamma \frac{\mathrm{d}\phi_1}{\mathrm{d}t} \hspace{2mm} ,\\
  \nonumber
  B_f(1+\gamma)\frac{\mathrm{d}m_2}{\mathrm{d}t}+B_f(1+\beta)\frac{\mathrm{d}m^f_2}{\mathrm{d}t}+B_f\delta\frac{\mathrm{d}m^s_2}{\mathrm{d}t}-\frac{A_s+B_s}{2}\alpha_1 e_s \frac{\mathrm{d}n^s_2}{\mathrm{d}t} \\
  \label{39}
  +\Omega_0(C_f+K_{CMB}A_f+K_{ICB}A_s)m^f_1-\Omega_0 K_{ICB}A_s m^s_1=B_f \gamma \frac{\mathrm{d}\phi_2}{\mathrm{d}t};
  \end{eqnarray}
  and combining equations (\ref{24}), (\ref{25}), (\ref{34}), and (\ref{35}), the rotational equation of the solid inner core then can be expressed as:
  \begin{eqnarray}
   \nonumber
   A_s(1+\theta)\frac{\mathrm{d}m_1}{\mathrm{d}t}+A_s\chi\frac{\mathrm{d}m^f_1}{\mathrm{d}t}+A_s(1+\nu)\frac{\mathrm{d}m^s_1}{\mathrm{d}t}+\frac{A_s+B_s}{2}e_s\frac{\mathrm{d}n^s_1}{\mathrm{d}t} \\
   \nonumber
   +\Omega_0 (C_s-B_s)(1-\alpha_1) m_2-[(C_s-B_s)\alpha_1-B_s K_{ICB}]\Omega_0 m^f_2-B_s(1+K_{ICB})\Omega_0 m^s_2 \\
   \label{40}
   -\Omega_0 n^s_2(\frac{A_s+B_s}{2}e_s-\alpha_2(C_s-B_s))=A_s\theta\frac{\mathrm{d}\phi_1}{\mathrm{d}t}+(C_s-B_s)\alpha_3\Omega_0\phi_2 \hspace{2mm} , \\
   \nonumber
   B_s(1+\theta)\frac{\mathrm{d}m_2}{\mathrm{d}t}+B_s\chi\frac{\mathrm{d}m^f_2}{\mathrm{d}t}+B_s(1+\nu)\frac{\mathrm{d}m^s_2}{\mathrm{d}t}+\frac{A_s+B_s}{2}e_s\frac{\mathrm{d}n^s_2}{\mathrm{d}t} \\
   \nonumber
   -\Omega_0 (C_s-A_s)(1-\alpha_1) m_1+[(C_s-A_s)\alpha_1-A_s K_{ICB}]\Omega_0 m^f_1+A_s(1+K_{ICB})\Omega_0 m^s_1 \\
   \label{41}
   +\Omega_0 n^s_1(\frac{A_s+B_s}{2}e_s-\alpha_2(C_s-A_s))=B_s\theta\frac{\mathrm{d}\phi_2}{\mathrm{d}t}-(C_s-A_s)\alpha_3\Omega_0\phi_1.
  \end{eqnarray}

The new rotational equations for the whole Earth, fluid outer core, and solid inner core, expressed by equations (\ref{36})-(\ref{41}), along with the tilt equations for the solid inner core, expressed by equations (\ref{27}) and (\ref{28}), constitute the basic equations of the triaxial three-layered Earth rotational theory, a special case where MBHS1991 theory \citep{mathews1991a,mathews2002} represents the rotationally symmetric case.

  %---------------------------------------------------------------------%
\section{Solution Formulation of the Rotational Normal Modes for Triaxial Three-layered Earth Rotation}
\label{S3}
In the rotationally symmetric three-layered Earth rotation case, \citet{mathews1991a} provided the analytical rotational normal mode solutions for the three-layered Earth with only pressure and gravitational couplings acting on the solid inner core. \citet{mathews2002} extended this model by incorporating the electromagnetic couplings near the CMB, ICB, and mantle anelasticity and ocean tide effects with the analytical rotational normal modes solutions as followings:
\begin{eqnarray}
\label{42}
\sigma_{CW}=\frac{A}{A_m}(e-\kappa) \\
\label{43}
\sigma_{FCN}=-1-(1+\frac{A_f}{A_m})(e_f-\beta+K_{CMB}+K_{ICB}\frac{A_s}{A_f}) \\
\label{44}
\sigma_{FICN}=-1+(1+\frac{A_s}{A_m})(\alpha_2 e_s+\nu-K_{ICB}) \\
\label{45}
\sigma_{ICW}=(1-\alpha_2)e_s=\alpha_3(\alpha_g+1)e_s,
\end{eqnarray}
where $e$, $e_f$ and $e_s$ are the dynamic ellipticities for the whole Earth, fluid outer core, and solid inner core, respectively. Here, $K_{CMB}$ and $K_{ICB}$ can be the electromagnetic coupling parameters or the viscoelectromagnetic coupling parameters depending on the model for dissipative coupling in the fluid outer core, i.e. the electromagnetic coupling model of \citet{mathews2002} or the viscoelectromagnetic coupling model of \citet{mathews2005}. The advantages of equations (\ref{42})-(\ref{45}) are that the influences of the compliances and core mantle coupling parameters on the normal modes can be clearly seen.

Regarding triaxial three-layered Earth rotation, it is more complex to solve for the rotational normal modes. In fact, there are two methods to solve the rotational normal modes for the triaxial three-layered Earth rotation equations, one was proposed by \citet{hoolst2002} and later used by \citet{chen2010}. The two methods are called the trigonometric function method and the eigenvalue method. They originated from the resonant frequency study of \citet{mathews2002} who considered the electromagnetic couplings near the CMB and ICB, mantle anelasticity, and ocean tides to study forced nutations, and developed by \citet{sun2015} to study the CW and FCN under the frame of the triaxial two-layered Earth rotation of \citet{chen2010}. In both methods, the external tidal potential terms should be set to zero for the rotational normal mode solutions. Here, we generalized these two methods to give the solution formulation of the rotational normal modes for the triaxial three-layered Earth rotation. 

\subsection{Trigometric function method}
\label{S3.1}
As suggested by \citet{hoolst2002} (see also \citet{chen2010}), the dimensionless rotational parameters for the whole Earth, the fluid outer core, the solid inner core, and the tilt of the solid inner core in the triaxial three-layered Earth rotation frame can be expressed in elliptic motions:
\begin{equation}
\begin{gathered}
\label{46}
m_1=m_{1c}\cos\sigma\Omega_0t+m_{1s}\sin\sigma\Omega_0t,\quad m_2=m_{2c}\cos\sigma\Omega_0t+m_{2s}\sin\sigma\Omega_0t \\
m^f_1=m^f_{1c}\cos\sigma\Omega_0t+m^f_{1s}\sin\sigma\Omega_0t,\quad m^f_2=m^f_{2c}\cos\sigma\Omega_0t+m^f_{2s}\sin\sigma\Omega_0t \\
m^s_1=m^s_{1c}\cos\sigma\Omega_0t+m^s_{1s}\sin\sigma\Omega_0t,\quad m^s_2=m^s_{2c}\cos\sigma\Omega_0t+m^s_{2s}\sin\sigma\Omega_0t \\
n^s_1=n^s_{1c}\cos\sigma\Omega_0t+n^s_{1s}\sin\sigma\Omega_0t,\quad n^s_2=n^s_{2c}\cos\sigma\Omega_0t+n^s_{2s}\sin\sigma\Omega_0t, 
\end{gathered}
\end{equation}
where $\sigma$ is the frequency of the rotational motions in cycles per sidereal day (cpsd). By substituting equation (\ref{46}) into equations (\ref{36})-(\ref{41}), (\ref{27}), and (\ref{28}), we obtained the normal modes equations expressed as
\begin{eqnarray}
\label{47}
%whole Earth
(1+\kappa)\sigma A m_{1s}+(A_f+A\xi)\sigma m^f_{1s}+(A_s+A\varsigma)\sigma m^s_{1s}+\frac{A_s+B_s}{2}\alpha_3 e_s\sigma n^s_{1s} \notag\\
+(C-(1+\kappa)B)m_{2c}-(B_f+B\xi)m^f_{2c}-(B_s+B\varsigma)m^s_{2c}-\frac{A_s+B_s}{2}\alpha_3 e_s n^s_{2c}=0 \notag\\
-(1+\kappa)\sigma A m_{1c}-(A_f+A\xi)\sigma m^f_{1c}-(A_s+A\varsigma)\sigma m^s_{1c}-\frac{A_s+B_s}{2}\alpha_3 e_s\sigma n^s_{1c} \notag\\
+(C-(1+\kappa)B)m_{2s}-(B_f+B\xi)m^f_{2s}-(B_s+B\varsigma)m^s_{2s}-\frac{A_s+B_s}{2}\alpha_3 e_s n^s_{2s}=0 \notag\\
(1+\kappa)\sigma B m_{2s}+(B_f+B\xi)\sigma m^f_{2s}+(B_s+B\varsigma)\sigma m^s_{2s}+\frac{A_s+B_s}{2}\alpha_3 e_s\sigma n^s_{2s} \notag\\
-(C-(1+\kappa)A)m_{1c}+(A_f+A\xi)m^f_{1c}+(A_s+A\varsigma)m^s_{1c}+\frac{A_s+B_s}{2}\alpha_3 e_s n^s_{1c}=0 \notag\\
-(1+\kappa)\sigma B m_{2c}-(B_f+B\xi)\sigma m^f_{2c}-(B_s+B\varsigma)\sigma m^s_{2c}-\frac{A_s+B_s}{2}\alpha_3 e_s\sigma n^s_{2c} \notag\\
-(C-(1+\kappa)A)m_{1s}+(A_f+A\xi)m^f_{1s}+(A_s+A\varsigma)m^s_{1s}+\frac{A_s+B_s}{2}\alpha_3 e_s n^s_{1s}=0 \notag\\
%fluid outer core
A_f(1+\gamma)\sigma m_{1s}+A_f(1+\beta)\sigma m^f_{1s}+A_f\delta\sigma m^s_{1s}-\frac{A_s+B_s}{2}\alpha_1 e_s \sigma n^s_{1s} \notag\\
-(C_f+K_{CMB}B_f+K_{ICB}B_s)m^f_{2c}+K_{ICB}B_s m^s_{2c}=0 \notag\\
-A_f(1+\gamma)\sigma m_{1c}-A_f(1+\beta)\sigma m^f_{1c}-A_f\delta\sigma m^s_{1c}+\frac{A_s+B_s}{2}\alpha_1 e_s \sigma n^s_{1c} \notag\\
-(C_f+K_{CMB}B_f+K_{ICB}B_s)m^f_{2s}+K_{ICB}B_s m^s_{2s}=0 \notag\\
B_f(1+\gamma)\sigma m_{2s}+B_f(1+\beta)\sigma m^f_{2s}+B_f\delta\sigma m^s_{2s}-\frac{A_s+B_s}{2}\alpha_1 e_s \sigma n^s_{2s} \notag\\
+(C_f+K_{CMB}A_f+K_{ICB}A_s)m^f_{1c}-K_{ICB}A_s m^s_{1c}=0 \notag\\
-B_f(1+\gamma)\sigma m_{2c}-B_f(1+\beta)\sigma m^f_{2c}-B_f\delta\sigma m^s_{2c}+\frac{A_s+B_s}{2}\alpha_1 e_s \sigma n^s_{2c} \notag\\
+(C_f+K_{CMB}A_f+K_{ICB}A_s)m^f_{1s}-K_{ICB}A_s m^s_{1s}=0 \\
%solid inner core
A_s(1+\theta)\sigma m_{1s}+A_s\chi\sigma m^f_{1s}+A_s(1+\nu)\sigma m^s_{1s} \notag\\
+\frac{A_s+B_s}{2}e_s\sigma n^s_{1s}+(C_s-B_s)(1-\alpha_1) m_{2c}-[(C_s-B_s)\alpha_1-B_s K_{ICB}] m^f_{2c} \notag\\
-B_s(1+K_{ICB})m^s_{2c}-(\frac{A_s+B_s}{2}e_s-\alpha_2(C_s-B_s))n^s_{2c}=0 \notag\\
-A_s(1+\theta)\sigma m_{1c}-A_s\chi\sigma m^f_{1c}-A_s(1+\nu)\sigma m^s_{1c} \notag\\
-\frac{A_s+B_s}{2}e_s\sigma n^s_{1c}+(C_s-B_s)(1-\alpha_1) m_{2s}-[(C_s-B_s)\alpha_1-B_s K_{ICB}] m^f_{2s} \notag\\
-B_s(1+K_{ICB})m^s_{2s}-(\frac{A_s+B_s}{2}e_s-\alpha_2(C_s-B_s))n^s_{2s}=0 \notag\\
B_s(1+\theta)\sigma m_{2s}+B_s\chi\sigma m^f_{2s}+B_s(1+\nu)\sigma m^s_{2s} \notag\\
+\frac{A_s+B_s}{2}e_s\sigma n^s_{2s}-(C_s-A_s)(1-\alpha_1) m_{1c}+[(C_s-A_s)\alpha_1-A_s K_{ICB}] m^f_{1c} \notag\\
+A_s(1+K_{ICB})m^s_{1c}+(\frac{A_s+B_s}{2}e_s-\alpha_2(C_s-A_s))n^s_{1c}=0 \notag\\
-B_s(1+\theta)\sigma m_{2c}-B_s\chi\sigma m^f_{2c}-B_s(1+\nu)\sigma m^s_{2c} \notag\\
-\frac{A_s+B_s}{2}e_s\sigma n^s_{2c}-(C_s-A_s)(1-\alpha_1) m_{1s}+[(C_s-A_s)\alpha_1-A_s K_{ICB}] m^f_{1s} \notag\\
+A_s(1+K_{ICB})m^s_{1s}+(\frac{A_s+B_s}{2}e_s-\alpha_2(C_s-A_s))n^s_{1s}=0 \notag\\
%inner core tilt
\sigma n^s_{1s}-m^s_{2c}=0 \notag\\
-\sigma n^s_{1c}-m^s_{2s}=0 \notag\\
\sigma n^s_{2s}+m^s_{1c}=0 \notag\\
-\sigma n^s_{2c}+m^s_{1s}=0, \notag
\end{eqnarray}
that can be rewritten in a matrix form as:
\begin{equation}
   \label{48}
   \bm{M}\cdot\bm{X}=0,
\end{equation}
where $\bm{M}$ is a $16\times16$ matrix, expressed as,
\begin{equation}
   \label{49}
   \bm{M}=[M_{ij}],\quad i,j=1,2,3,4,
\end{equation}
where $M_{ij}$ are a set of $4\times4$ matrixes, expressed in {\textbf{Class I} of Appendix A, and $\bm{X}$ is a $16\times1$ column vector written as
\begin{eqnarray}
 \label{50}
 \bm{X}^T=[m_{1c}\ m_{1s}\ m_{2c}\ m_{2s}\ m^f_{1c}\ m^f_{1s}\ m^f_{2c}\ m^f_{2s}\ m^s_{1c}\ m^s_{1s}\ m^s_{2c}\ m^s_{2s}\ n^s_{1c}\ n^s_{1s}\ n^s_{2c}\ n^s_{2s}].
\end{eqnarray}
The rotational normal modes of the triaxial three-layered Earth can be solved by setting $|\bm{M}|=0$. 

\subsection{Eigenvalue Method}
\label{S3.2}
Generalizing the eigenvalue method proposed by \citet{sun2015} in the frame of a triaxial two-layered Earth rotation, we formulated a triaxial three-layered Earth rotation theory in the form of:
\begin{eqnarray}
\label{51}
(1+\kappa)A\frac{\mathrm{d}m_1}{\mathrm{d}t}+(A_f+A\xi)\frac{\mathrm{d}m^f_1}{\mathrm{d}t}+(A_s+A\varsigma)\frac{\mathrm{d}m^s_1}{\mathrm{d}t}+\frac{A_s+B_s}{2}\alpha_3 e_s \frac{\mathrm{d}n^s_1}{\mathrm{d}t} \notag\\
    =-(C-(1+\kappa)B)\Omega_0 m_2+(B_f+B\xi)\Omega_0 m^f_2+(B_s+B\varsigma)\Omega_0 m^s_2+\frac{A_s+B_s}{2}\alpha_3 e_s \Omega_0 n^s_2 \notag\\
(1+\kappa)B\frac{\mathrm{d}m_2}{\mathrm{d}t}+(B_f+B\xi)\frac{\mathrm{d}m^f_2}{\mathrm{d}t}+(B_s+B\varsigma)\frac{\mathrm{d}m^s_2}{\mathrm{d}t}+\frac{A_s+B_s}{2}\alpha_3 e_s \frac{\mathrm{d}n^s_2}{\mathrm{d}t} \notag\\
    =(C-(1+\kappa)A)\Omega_0 m_1-(A_f+A\xi)\Omega_0 m^f_1-(A_s+A\varsigma)\Omega_0 m^s_1-\frac{A_s+B_s}{2}\alpha_3 e_s \Omega_0 n^s_1 \notag\\
A_f(1+\gamma)\frac{\mathrm{d}m_1}{\mathrm{d}t}+A_f(1+\beta)\frac{\mathrm{d}m^f_1}{\mathrm{d}t}+A_f\delta\frac{\mathrm{d}m^s_1}{\mathrm{d}t}-\frac{A_s+B_s}{2}\alpha_1 e_s \frac{\mathrm{d}n^s_1}{\mathrm{d}t} \notag\\
  =\Omega_0(C_f+K_{CMB}B_f+K_{ICB}B_s)m^f_2-\Omega_0 K_{ICB}B_s m^s_2 \notag\\
B_f(1+\gamma)\frac{\mathrm{d}m_2}{\mathrm{d}t}+B_f(1+\beta)\frac{\mathrm{d}m^f_2}{\mathrm{d}t}+B_f\delta\frac{\mathrm{d}m^s_2}{\mathrm{d}t}-\frac{A_s+B_s}{2}\alpha_1 e_s \frac{\mathrm{d}n^s_2}{\mathrm{d}t} \notag\\
  =-\Omega_0(C_f+K_{CMB}A_f+K_{ICB}A_s)m^f_1+\Omega_0 K_{ICB}A_s m^s_1 \notag\\
A_s(1+\theta)\frac{\mathrm{d}m_1}{\mathrm{d}t}+A_s\chi\frac{\mathrm{d}m^f_1}{\mathrm{d}t}+A_s(1+\nu)\frac{\mathrm{d}m^s_1}{\mathrm{d}t}+\frac{A_s+B_s}{2}e_s\frac{\mathrm{d}n^s_1}{\mathrm{d}t} \\
  =-\Omega_0 (C_s-B_s)(1-\alpha_1) m_2+[(C_s-B_s)\alpha_1-B_s K_{ICB}]\Omega_0 m^f_2 \notag\\
  +B_s(1+K_{ICB})\Omega_0 m^s_2 
   +\Omega_0 n^s_2(\frac{A_s+B_s}{2}e_s-\alpha_2(C_s-B_s)) \notag\\
B_s(1+\theta)\frac{\mathrm{d}m_2}{\mathrm{d}t}+B_s\chi\frac{\mathrm{d}m^f_2}{\mathrm{d}t}+B_s(1+\nu)\frac{\mathrm{d}m^s_2}{\mathrm{d}t}+\frac{A_s+B_s}{2}e_s\frac{\mathrm{d}n^s_2}{\mathrm{d}t} \notag\\
   =\Omega_0 (C_s-A_s)(1-\alpha_1) m_1-[(C_s-A_s)\alpha_1-A_s K_{ICB}]\Omega_0 m^f_1 \notag\\
   -A_s(1+K_{ICB})\Omega_0 m^s_1 
   -\Omega_0 n^s_1(\frac{A_s+B_s}{2}e_s-\alpha_2(C_s-A_s)) \notag\\
\frac{\mathrm{d}n^s_1}{\mathrm{d}t}=\Omega_0 m^s_2 \notag\\
\frac{\mathrm{d}n^s_2}{\mathrm{d}t}=-\Omega_0 m^s_1, \notag        
\end{eqnarray}
which can be written into matrix form
\begin{equation}
\label{52}
\mathbf{F}\frac{\mathrm{d}\mathbf{y}}{\mathrm{d}t}=\Omega_0 \mathbf{G}\mathbf{y},
\end{equation}
where
\begin{equation}
\label{53}
\mathbf{y}=\big[m_1\ m_2\ m^f_1\ m^f_2\ m^s_1\ m^s_2\ n^s_1\ n^s_2\big]^T,
\end{equation}
and
\begin{equation*}
\mathbf{F}=\left[
\begin{array}{cccc}
A(1+\kappa)&0&A_f+A\xi&0 \\
0&B(1+\kappa)&0&B_f+B\xi \\
A_f(1+\gamma)&0&A_f(1+\beta)&0 \\
0&B_f(1+\gamma)&0&B_f(1+\beta) \\
A_s(1+\theta)&0&A_s\chi&0 \\
0&B_s(1+\theta)&0&B_s\chi \\
0&0&0&0 \\
0&0&0&0
\end{array}\right.
\end{equation*}
%\begin{linenomath*}
\begin{equation}
\label{54}
\left.\begin{array}{cccc}
A_s+A\varsigma&0&(A_s+B_s)\alpha_3e_s/2&0 \\
0&B_s+B\varsigma&0&(A_s+B_s)\alpha_3e_s/2 \\
A_f\delta&0&-(A_s+B_s)\alpha_3e_s/2&0 \\
0&B_f\delta&0&-(A_s+B_s)\alpha_3e_s/2 \\
A_s(1+\nu)&0&(A_s+B_s)e_s/2&0 \\
0&B_s(1+\nu)&0&(A_s+B_s)e_s/2 \\
0&0&1&0 \\
0&0&0&1
\end{array}
\right]
\end{equation}
and
\begin{equation*}
\mathbf{G}=\left[
\begin{array}{cccc}
0&-(C-(1+\kappa)B)&0&B_f+B\xi \\
C-(1+\kappa)A&0&-(A_f+A\xi)&0 \\
0&0&0&C_f+K_{CMB}B_f+K_{ICB}B_s \\
0&0&-(C_f+K_{CMB}A_f+K_{ICB}A_s)&0 \\
0&-(C_s-B_s)(1-\alpha_1)&0&(C_s-B_s)\alpha_1-B_s K_{ICB} \\
(C_s-A_s)(1-\alpha_1)&0&-[(C_s-A_s)\alpha_1-A_s K_{ICB}]&0 \\
0&0&0&0 \\
0&0&0&0
\end{array}\right.
\end{equation*}
\begin{equation}
\label{55}
\left.\begin{array}{cccc}
0&B_s+B\varsigma&0&(A_s+B_s)\alpha_3 e_s/2 \\
-(A_s+A\varsigma)&0&-(A_s+B_s)\alpha_3 e_s/2&0 \\
0&-K_{ICB}B_s&0&0 \\
K_{ICB}A_s&0&0&0 \\
0&B_s(1+K_{ICB})&0&(A_s+B_s)e_s/2-\alpha_2(C_s-B_s) \\
-A_s(1+K_{ICB})&0&-[(A_s+B_s)e_s/2-\alpha_2(C_s-A_s)]&0 \\
0&1&0&0 \\
-1&0&0&0
\end{array}
\right].
\end{equation}
According to the theory of ordinary differential equations, the solution of an equation system (\ref{51}) has the form $\mathbf{r}_i e^{\sigma_i t}$, where $\sigma_i$ are the eigenvalues of $\mathbf{F}^{-1}\mathbf{G}$, because the motions of $\mathbf{y}$ are in fact sinusoidal and hence $-i\sigma_i$ are the rotational normal mode solutions used in this study since $\sigma_i=i(-i\sigma_i)$.

\section{Model Input Data}
\label{S4}
In normal mode calculations, the needed input data are the dynamical figure parameters of the triaxial three-layered Earth, including the principal moments of inertia and the dynamic ellipticities of the triaxial solid mantle, triaxial fluid outer core, and triaxial solid inner core. In addition, the compliances due to tidal deformation and non-uniform rotational deformation, along with the core mantle coupling parameters, are also needed. The core mantle coupling parameters include the viscoelectromagnetic coupling parameters and the pressure and gravitational coupling parameters. 

\begin{table}[htp]
  \centering
  \caption{Dynamical figure parameters, compliances, and core mantle coupling parameters for the triaxial three-layered Earth (Note that $e$, $e_f$, and $e_s$ in the table are dimensionless)$^*$}
  \label{T1}
  \begin{tabular}{cccccc}
  \specialrule{0.1em}{0.5pt}{0.5pt}
  \multicolumn{1}{c}{\multirow{2}*{}} & {value}  & {\multirow{2}*{compliances}} &  {\multirow{2}*{value}} &  {core mantle coupling} & {\multirow{2}*{value}}   \\
  \multicolumn{1}{c}{} & {($\mathrm{kg\ m^2}$)} & {} & {} & {parameters} & {} \\  
  \specialrule{0.05em}{0.25pt}{0.25pt}
  \multicolumn{1}{c}{\multirow{2}*{$A$}} &  {\multirow{2}*{$8.0085082\times10^{37}$}}  &   {\multirow{2}*{$\kappa$}}  &  {$1.242\times10^{-3}$}  & {\multirow{2}*{$K_{CMB}$}} &   {$2.97\times10^{-5}$}             \\
  \multicolumn{1}{c}{}  & {} & {} & {$-1.195\times10^{-5}i$} & {} & {$-1.78\times10^{-5}i$} \\
  \multicolumn{1}{c}{\multirow{2}*{$B$}}  & {\multirow{2}*{$8.0086847\times10^{37}$}} & {\multirow{2}*{$\xi$}} & {\multirow{2}*{$2.460\times10^{-4}$}} & 
  {\multirow{2}*{$K_{ICB}$}} & {$1.01\times10^{-3}$}          \\
  \multicolumn{1}{c}{} & {} & {} & {} & {} & {$-1.09\times10^{-3}i$} \\
  $C$    &  $8.0349010\times10^{37}$  & $\varsigma$ &  $4.964\times10^{-9}$                      &       $\alpha_1$     &   0.9463         \\
  \multicolumn{1}{c}{\multirow{2}*{$A_f$}}  &  {\multirow{2}*{$9.0549367\times10^{36}$}}  &   {\multirow{2}*{$\gamma$}}  &  {$1.964\times10^{-3}$}  &  {\multirow{2}*{$\alpha_2$}}     &   {\multirow{2}*{0.829492}}         \\
  \multicolumn{1}{c}{} & {} & {} & {$-1.073\times10^{-4}i$} & {} & {} \\
  $B_f$  &  $9.0550794\times10^{36}$  &   $\beta$   &  $6.262\times10^{-4}$                      &       $\alpha_3$     &   0.0537         \\
  $C_f$  &  $9.0789730\times10^{36}$  &   $\delta$  &  $-4.869\times10^{-7}$                     &       $\alpha_g$     &   2.1752     \\
  $A_s$  &  $5.8509312\times10^{34}$  &   $\theta$  &  $6.794\times10^{-6}$                      &                      &         \\
  $B_s$  &  $5.8510262\times10^{34}$  &   $\chi$    &  $-7.536\times10^{-5}$                     &                      &         \\
  $C_s$  &  $5.8651498\times10^{34}$  &   $\nu$     &  $7.984\times10^{-5}$                      &                      &         \\
   $e$   &  $0.0032845479$            &             &                                            &                      &         \\
  $e_f$  &  $0.0026456000$            &             &                                            &                      &         \\
  $e_s$  &  $0.0024220000$            &             &                                            &                      &         \\
  \specialrule{0.1em}{0.25pt}{0.25pt}
  \end{tabular}
    \vspace{2mm}
    \begin{flushleft}
  \small $^*$ Here, the dynamical figure parameters were obtained from \citet{chen2015} after extending the method of \citet{chen2010}. The compliances were from \citet{yang2016} and \citet{mathews2002}. The viscoelectromagnetic coupling parameters were from \citet{koot2010}, while the pressure and gravitational coupling parameters were from \citet{mathews1991b}, from which $\alpha_2$ was recalculated according to the relation $\alpha_2=\alpha_1-\alpha_3\alpha_g$.
    \end{flushleft}
\end{table}

The dynamical figure parameters for the triaxial three-layered Earth were from \citet{chen2015}, see Table \ref{T1}, where $e$ and $e_f$ were from \citet{mathews2002} and were determined by fitting the theoretical nutation amplitudes to the precession rate and nutation data, while $e_s$ is a theoretical value from the PREM model, and the principal moments of inertia for the whole Earth were obtained using second-order gravitational potential coefficients and the dynamic ellipticity of the Earth based on an eigenvalue-eigenvector approach \citep{marchenko2001,marchenko2003}. Combining the MHB2000 Earth model and reasonable assumptions about the Earth's internal figure, the principal moments of inertia of the fluid outer core and solid inner core were obtained after extending the method of \citet{chen2010}.

 Listed also in Table \ref{T1} are the values of the compliances $\kappa$, $\xi$, $\gamma$, and $\beta$ from the study of \citet{yang2016}, and the remaining compliances are from \citet{mathews2002}. The compliances $\kappa$ and $\gamma$ are complex due to mantle anelasticity and ocean tide dissipations. The parameter $K_{CMB}$ represents the coupling strength of the viscoelectromagnetic coupling torque acting on the fluid outer core by the mantle, while the parameter $K_{ICB}$ represents the coupling strength of the viscoelectromagnetic coupling torque acting on the solid inner core by the fluid outer core. The $K_{CMB}$ and $K_{ICB}$ values were obtained from fitting the results of the nutation observations by \citet{koot2010}, both of which are dimensionless complex numbers. The pressure and gravitational coupling parameters $\alpha_1$ and $\alpha_g$ were calculated from the PREM model by \citet{mathews1991b}. The relevant values are listed in Table \ref{T1}. 

%-------------------------------------------------------------------%
\section{Results of Rotational Normal Mode Solutions}
\label{S5}
Using the parameters listed in Table \ref{T1}, the rotational normal modes of the triaxial three-layered Earth rotation equations can be solved numerically using the trigometric function or eigenvalue methods. Here, we provide the rotational normal mode solutions for eight different cases. Both the trigometric function and eigenvalue methods will generate the same results except in Case IV that had a very slight difference in the FICN quality factor. Cases I and II correspond to the triaxial three-layered case and rotationally symmetric three-layered case, respectively, with both viscoelectromagnetic couplings and pressure and gravitational couplings considered. Cases III and IV were designed to investigate the dissipative properties of the solid Earth reflected from normal mode solutions and will be discussed in the next section. In Case V, the normal modes were fitted to the mean of the CW observations (see Table \ref{T2}) and to the mean of the FCN observations (see Table \ref{T3}). Case VI corresponds to the normal mode solutions in a rotationally symmetric case with the newly constrained compliances. In Case VII, the viscous couplings were not considered, and we deployed the electromagnetic coupling parameters in \citet{mathews2002} to calculate the rotational normal mode solutions for comparison, i.e. $K_{CMB}=2.245\times10^{-5}-1.85\times10^{-5}i$ and $K_{ICB}=1.11\times10^{-3}-0.78\times10^{-3}i$. In Case VIII, we also only considered the electromagnetic couplings and obtained the rotational normal mode solutions for the rotationally symmetric case. The normal mode solutions for these eight cases are provided in Table \ref{T4}. 

By comparing the normal mode solutions from Cases I and II with the viscoelectromagnetic couplings and Cases VII and VIII with only the electromagnetic couplings, we found that the triaxiality will prolong the CW and ICW by about 0.01 d and 0.35 d, respectively. By comparing the normal mode solutions from Case II (rotationally symmetric case under the triaxial three-layered Earth rotation theory frame) with the normal mode solutions of the MBHS1991 theory, we found that the two solutions were the same, and hence the normal mode results of MBHS1991 theory represents special cases to our general solutions. 

\begin{table}[htp]
\centering
\caption{The CW observations cited from Table 2 of \citet{nastula2015}}
\label{T2}
\begin{tabular}{cc}
\specialrule{0.1em}{0.5pt}{0.5pt}
       CW          &    Reference \\
\specialrule{0.05em}{0.25pt}{0.25pt}
    433.2/63       & \citet{jeffreys1972}   \\
   434.0/100       & \citet{wilson1976}     \\
    434.8/96       & \citet{ooe1978}        \\
   433.3/170       & \citet{wilson1980}     \\
   433.0/179       & \citet{wilson1990}     \\
    439.5/72       & \citet{kuehne1996}     \\
    433.7/49       & \citet{furuya1996}     \\
    430.8/41       & \citet{gross2005}      \\
   429.4/107       & \citet{gross2005}      \\
    431.9/83       & \citet{gross2005}      \\
   432.98/97       & \citet{seitz2012}      \\
  430.3/88.4       & \citet{mathews2002}    \\
  433.03/100.20    & \citet{chen2010}       \\
   430.9/127       & \citet{nastula2015}    \\
\specialrule{0.1em}{0.25pt}{0.25pt}
\end{tabular}
\end{table}

\begin{table}[htp]
\centering
\caption{The FCN observations cited from Table 1 of \citet{rosat2009b}}
\label{T3}
\begin{tabular}{cc}
\specialrule{0.1em}{0.5pt}{0.5pt}
       FCN                &    Reference \\
\specialrule{0.05em}{0.25pt}{0.25pt}
    431/2800              &  \citet{neuberg1987}      \\
  435/$22000-10^5$        &  \citet{herring1986}      \\
  428/3300-37000          &  \citet{cummins1993}      \\
    437/3200              &  \citet{sato1994}         \\
  424/2300-8300           &  \citet{defraigne1994}\,(Stacked gravity)        \\
    432/>15000            &  \citet{defraigne1994}\,(VLBI)    \\
    433/>17000            &  \citet{defraigne1994}\,(Stacked gravity+VLBI)   \\
  431/>1700-2500          &  \citet{florsch1994}      \\ 
  430/5500-10000          &  \citet{merriam1994}      \\
      429/7700            &  \citet{hinderer1995}     \\
    428/>20000            &  \citet{florsch2000}      \\
  430.20/20000            &  \citet{mathews2002}      \\
 429.7/9350-10835         &  \citet{sato2004}         \\
     430/17000            &  \citet{lambert2007}      \\
     430/15000            &  \citet{ducarme2009}      \\ 
     430/13750            &  \citet{koot2008}         \\
 428/7762-31989           &  \citet{rosat2009b}       \\
\specialrule{0.1em}{0.25pt}{0.25pt}
\end{tabular}
\end{table}

\begin{table}[htp]
\centering
\caption{Normal mode solutions for various cases of the Triaxial Three-Layered Earth Rotation Theory: Case I, triaxial case; Case II, rotationally symmetric case; Case III, without mantle anelasticity and ocean tide dissipations($Im(\kappa)=Im(\gamma)=0$, $Im(K_{CMB},K_{ICB})\neq 0$); Case IV, without viscoelectromagnetic couplings induced dissipations($Im(\kappa,\gamma)\neq 0$, $Im(K_{CMB})=Im(K_{ICB})=0)$); Case V, fit to observations from the triaxial case; Case VI, normal mode solutions from the rotationally symmetric case with the constrained compliances; Case VII, normal mode solutions with the electromagnetic coupling parameters from \citet{mathews2002}, where the viscous couplings near CMB and ICB were not considered; Case VIII, normal mode solutions from the rotationally symmetric case with only the electromagnetic couplings from \citet{mathews2002} considered}
\label{T4}
\begin{tabular}{cccccc}
\specialrule{0.1em}{0.5pt}{0.5pt}
                                            &             &       CW      &       FCN      &       FICN      &         ICW        \\
\specialrule{0.05em}{0.25pt}{0.25pt}
\multicolumn{1}{c}{\multirow{2}*{Case I}}   &  {Period}   &    {433.19}   &    {-429.86}    &    {921.69}    &     {2413.01}        \\
\multicolumn{1}{c}{}                        &     {Q}     &     {85.44}   &   {22737.21}   &     {456.31}    &      {458.19}        \\
\multicolumn{1}{c}{\multirow{2}*{Case II}}  &  {Period}   &    {433.18}   &    {-429.86}    &    {921.69}    &     {2412.66}        \\
\multicolumn{1}{c}{}                        &     {Q}     &     {85.44}   &   {22738.35}   &     {456.31}    &      {458.20}        \\
\multicolumn{1}{c}{\multirow{2}*{Case III}} &  {Period}   &    {433.19}   &    {-429.86}    &    {921.69}    &     {2413.01}        \\
\multicolumn{1}{c}{}                        &     {Q}     &  {219259.06}  &   {22770.06}   &     {456.31}    &      {458.45}        \\
\multicolumn{1}{c}{\multirow{2}*{Case IV}}  &  {Period}   &    {433.19}   &    {-429.86}    &    {1015.55}    &     {2413.25}        \\
\multicolumn{1}{c}{}                        &     {Q}     &     {85.47}   & {15822391.00}  & {$2.737576\times10^{10}$} &  {830509.02}     \\
\multicolumn{1}{c}{\multirow{2}*{Case V}}   &  {Period}   &  {$\mathbf{432.92}$}  &   {$\mathbf{-429.86}$}    &    {921.69}   &     {2413.01}        \\
\multicolumn{1}{c}{}                        &     {Q}     &         {85.49}       &        {22735.56}        &     {456.31}   &      {458.19}        \\
\multicolumn{1}{c}{\multirow{2}*{Case VI}}  &  {Period}   &        {432.91}       &         {-429.86}         &    {921.69}   &     {2412.66}        \\
\multicolumn{1}{c}{}                        &     {Q}     &         {85.50}       &        {22736.70}        &     {456.31}   &      {458.20}        \\
\multicolumn{1}{c}{\multirow{2}*{Case VII}} &  {Period}   &    {433.19}   &    {-431.72}    &    {1015.55}    &     {2413.25}        \\
\multicolumn{1}{c}{}                        &     {Q}     &     {85.44}   &   {22353.43}   &     {637.85}    &      {640.22}        \\
\multicolumn{1}{c}{\multirow{2}*{Case VIII}}  &  {Period}   &    {433.18}   &    {-431.72}    &    {1015.55}    &     {2412.90}        \\
\multicolumn{1}{c}{}                        &     {Q}     &     {85.44}   &   {22353.93}   &     {637.85}    &      {640.22}        \\
\specialrule{0.1em}{0.25pt}{0.25pt}
\end{tabular}
\end{table}

\subsection{Dissipative Properties of the Normal Modes}
\label{S5.1}
Generally, there are three kinds of dissipative processes in rotational normal modes \citep{smith1981}; mantle anelasticity induced dissipation, ocean tide induced dissipation, and viscoelectromagnetic coupling induced dissipation. Here, only the compliances $\kappa$, $\gamma$ and the two viscoelectromagnetic coupling parameters $K_{CMB}$, $K_{ICB}$ are complex numbers due to the dissipation caused by mantle anelasticity and ocean tides and that caused by viscoelectromagnetic coupling processes, respectively (\citet{mathews2002}). In this section, the three kinds of dissipative processes affecting the rotational normal modes under the frame of the triaxial three-layered Earth rotation theory and the calculation procedures will be discussed, focusing on the dissipations-sensitive quality factors of the normal modes rather than the periods variations.

Suppose there were no mantle anelasticity and ocean tide induced dissipations, and only viscoelectromagnetic coupling induced dissipations existed. For Case III, this means that the imaginary components of the compliances $\kappa$, $\gamma$ were set to zero (i.e.\,$Im(\kappa)=Im(\gamma)=0$, and $Im(K_{CMB},K_{ICB})\neq 0$). In this case, the rotational normal modes were solved, as listed in Table \ref{T4}. From the Case III results reported in Table \ref{T4}, it can be observed that the quality factor $Q$ of the CW became very large, while the FCN quality factor $Q$ varied slightly, and the other two FICN and ICW rotational normal modes quality factors remained nearly unchanged. Therefore, we can conclude that the mantle anelasticity and ocean tide induced dissipations dominate the CW motion, and this caused the CW quality factor $Q$ to change from a large value $219259.06$ to $85.44$, while they contributed slightly to the FCN motion.

Suppose there were no viscoelectromagnetic coupling induced dissipations, and there only the mantle anelasticity and ocean tide induced dissipations existed. For Case IV, this means that the imaginary components of the viscoelectromagnetic coupling parameters $K_{CMB}$, $K_{ICB}$ were set to zero (i.e.\,$Im(\kappa,\gamma)\neq 0$, and $Im(K_{CMB})=Im(K_{ICB})=0$). In this case, the rotational normal modes were solved and the results are listed in Table \ref{T4}. It can be observed that the CW quality factor $Q$ remained  nearly unchanged, while the $Q$ quality factors for the rotational normal modes FCN, FICN, and ICW varied significantly. Hence, we conclude that the viscoelectromagnetic coupling induced dissipations are responsible for the rotational normal modes FCN, FICN, and ICW, with almost no effect on the CW normal mode.

\subsection{Fitting to the Observations}
\label{S5.2}
There are quite a few observational results for the CW and FCN. For instance, \citet{nastula2015} estimated the CW period as 430.9 d and the $Q$ value as 127 by minimizing the modeled and observed polar motion excitation function (see Table \ref{T2}). \citet{rosat2009b} analyzed the superconducting gravimeter data in Europe using the resonance method and obtained the FCN period as 428 d with a quality factor of $7762<Q<31989$\ (see Table \ref{T3}). Here, we used the mean of the CW observations, $432.915$ d, and the mean of the FCN observations, $430.347$ d, to invert the compliances by fitting the calculated normal mode solutions to these observations. There is a FICN possible resonance effect with a period of 1300 sidereal days at the diurnal prograde frequency band, as revealed by \citet{rosat2016} using superconducting gravimeter data and VLBI celestial pole offset series that has not been verified. Thus, corresponding to Case V, we only fit the CW and FCN observations to constrain the compliances in the triaxial case. It can be seen from the analytical formulas of the rotational normal modes (\ref{42}) and (\ref{43}) that, if $\kappa$ becomes larger, then the CW period becomes longer, and if $\beta$ becomes larger, then the FCN period becomes longer. The viscoelectromagnetic coupling parameters were fitted from the nutation data by \citet{koot2010} and could be treated as constants, since the viscoelectromagnetic couplings were not very strong and only affect the CW, FCN, and ICW slightly, while the role of the viscoelectromagnetic couplings in the FICN will be carefully investigated in next section. The pressure and gravitational coupling parameters were determined by the structure and density of the Earth based on the PREM model or 1066A model \citep{mathews1991b} and can be regarded as constants.

Hence in Case V, if we set the CW observation as $432.92$ d and the FCN observation as $430.35$ d, then the inverted values of compliances were $\kappa=1.24063\times10^{-3}-1.195\times10^{-5}i$, $\gamma=1.601\times10^{-3}-1.073\times10^{-4}i$, and $\beta=6.238\times10^{-4}$, with other compliances unchanged, while the corresponding CW period and FCN period were fitted to $432.91$ d and $429.86$ d, respectively. During the fitting process, the FCN period was not sensitive to the $\beta$ parameter and the fitting error was about 0.5 d. The normal mode solutions for Case V are shown in Table \ref{T4}. If we used the new inverted compliances and other necessary data to calculate the normal mode solutions in the rotationally symmetric case, namely Case VI, the results showed that the CW and ICW periods were respectively prolonged by about 0.01 and 0.35 d after considering the triaxiality, while the triaxiality had no effect on the FCN and FICN in current parameter settings. The normal mode solutions for Case VI are also shown in Table \ref{T4}. 

\section{Additional Investigations on the Rotational Normal Modes}
\label{S6}
In section \ref{S5}, we mainly focused on the normal mode solutions in the triaxial and rotational symmetric cases with the solutions also fitted from observations, and on the Earth dissipation properties of the rotational normal modes. In the next section, we investigate the core mantle coupling effects on the rotational normal modes of the triaxial three-layered anelastic Earth rotation model in section \ref{S6.1}, and compare the rotational normal mode solutions with those of the rotationally symmetric three-layered non-rigid Earth model formulated by \citet{escapa2001}. Additionally, we compared the rotational normal mode solutions of the triaxial two-layered Earth rotation model formulated by \citet{chen2010} with the solutions of \citet{chen2010} in section \ref{S6.2}. Here, we used both the trigometric function and eigenvalue methods, but only provide the results from the trigometric function method illustrating the very slight differences.

\subsection{Core mantle coupling effects}
\label{S6.1}
The triaxial three-layered anelastic Earth model consists of the anelastic mantle, fluid outer core, and solid inner core. According to the core-mantle coupling models in the rotation model, we calculated the rotational normal modes corresponding to the following eight cases:\\
{\bf Case I}. The electromagnetic couplings, pressure, and gravitational couplings were considered. We used the values of the viscoelectromagnetic coupling parameters as given by \citet{koot2010}, and the pressure and gravitational coupling parameters as given by \citet{mathews1991b}.\\
{\bf Case II}. Only pressure and gravitational couplings were considered, meaning that the viscoelectromagnetic coupling parameters $K_{CMB}$ and $K_{ICB}$ were set to zero.\\
{\bf Case III}. Only the viscoelectromagnetic couplings were considered. Note that the pressure and gravitational coupling torque is expressed as equation (\ref{8}), and in this case we needed to set the pressure and gravitational coupling torque to zero to re-derive the motion equation of the solid inner core (equation (\ref{4})) and thus obtained the new normal mode matrix.\\
{\bf Case IV}. No couplings were considered.\\
{\bf Case V}. Only the pressure coupling was considered. By carefully examining the pressure and gravitational coupling torque given by \citet{mathews1991a}, we separated the pressure coupling torque from the gravitational coupling torque. The pressure coupling torque is expressed as:
\begin{equation}
\label{56}
\bm{\Gamma}_p=\Omega^2_0 \left( \begin{array}{c} 
\vspace{3mm}
(C_s-B_s)\big[\alpha_1 (m_2+m^f_2)-\alpha_1 n^s_2 - \alpha_1 \phi_2 \big] - c^s_{23} \\
\vspace{3mm}
-(C_s-A_s)\big[\alpha_1 (m_1+m^f_1)-\alpha_1 n^s_1 - \alpha_1 \phi_1 \big] + c^s_{13} \\
0
\end{array}\right).
\end{equation}
We replaced the original pressure and gravitational coupling torque $\bm{\Gamma}_s$ with this pressure coupling torque $\bm{\Gamma}_p$ and rederived the normal mode matrix. \\
{\bf Case VI}. Only the gravitational coupling was considered. As discussed in Case V, the gravitational coupling torque can be expressed as:
\begin{equation}
\label{57}
\bm{\Gamma}_g=\Omega^2_0 \left( \begin{array}{c} 
\vspace{3mm}
(C_s-B_s)\big[\alpha_g n^s_2 + \phi_2 \big] \\
\vspace{3mm}
-(C_s-A_s)\big[\alpha_g n^s_1 + \phi_1 \big] \\
0
\end{array}\right).
\end{equation}
where the terms associated with $n^s_1$ and $n^s_2$ are related to the gravitational coupling torque acting on the solid inner core by the fluid outer core and solid mantle. If they are combined with the pressure coupling, the gravitational coupling will be damped by a factor of $\alpha_3$ with the physical meaning of a density jump at the ICB. The terms associated with $\phi_i$ (i=1,2) are due to external tidal attraction. In this case, we replaced the original pressure and gravitational coupling torque $\bm{\Gamma}_s$ with this gravitational coupling torque $\bm{\Gamma}_g$ and rederived the normal mode matrix.\\
{\bf Case VII}. Consider only pressure coupling and viscoelectromagnetic coupling.  \\
{\bf Case VIII}. Only the gravitational coupling and viscoelectromagnetic coupling were considered. \\
For both the trigometric function and eigenvalue methods, we divided the different core mantle coupling conditions into four classes, and the corresponding normal mode matrixes and eigenvalue matrixes are present the results in Appendix \textbf{A} and \textbf{\ref{B}}.

Next, we used the parameters as given in section \ref{S4} to calculate the rotational normal modes for the eight core-mantle coupling cases using both the trigometric function and eigenvalue methods. Both methods generated approximately the same results, where the eigenvalue method led to very slight differences in the quality factors in some cases compared with trigometric function method. Hence, only the results corresponding to the eight coupling cases based on the trigometric function method are listed in Table \ref{T5}. We investigated the effect of pressure coupling, gravitational coupling, and electromagnetic couplings on the rotational normal modes by comparing the different cases.

\begin{table}[htp]
\centering
\caption{Rotational normal mode solutions for cases I, II, III, IV, V, VI, VII, and VIII for the triaxial three-layered anelastic Earth model: case I, both viscoelectromagnetic and pressure and gravitational couplings; case II, only pressure and gravitational couplings; case III, only viscoelectromagnetic couplings; case IV, no couplings; case V, only pressure coupling; case VI, only gravitational coupling; case VII, only viscoelectromagnetic and pressure couplings; VIII, only viscoelectromagnetic and gravitational couplings}
\label{T5}
\begin{tabular}{cccccc}
\specialrule{0.1em}{0.5pt}{0.5pt}
                                            &             &       CW       &       FCN      &       FICN      &         ICW        \\
 \specialrule{0.05em}{0.25pt}{0.25pt}
\multicolumn{1}{c}{\multirow{4}*{Case I}}   &  Frequency  & {0.00230359+}   &  {-1.00232+}    &  {-0.998918+}    &  {0.000413289+}      \\
\multicolumn{1}{c}{}                        &   (cpsd)    &{i0.0000134723}  & {i0.000022043} & {i0.00109455}   & {i$4.50997\times10^{-7}$} \\
\multicolumn{1}{c}{}                        &  {Period}   &    {432.92}    &    {-429.86}     &    {921.69}     &     {2413.01}        \\
\multicolumn{1}{c}{}                        &     {Q}     &     {85.49}     &   {22735.56}    &     {456.31}     &      {458.19}        \\
\multicolumn{1}{c}{\multirow{4}*{Case II}}  &  Frequency  & {0.0023036+}   &  {-1.00229+}     &  {-0.997903+}    &  {0.000413708+}      \\
\multicolumn{1}{c}{}                        &   (cpsd)    &{i0.0000134671}  & {i$3.13405\times10^{-8}$} & {i$4.03771\times10^{-11}$} & {i$2.48782\times10^{-10}$}    \\
\multicolumn{1}{c}{}                        &  {Period}   &    {432.918}    &    {-435.49}     &      {475.57}    &     {2410.56}        \\
\multicolumn{1}{c}{}                        &     {Q}     &     {85.53}     & {15990332.00}   & {$1.2357289\times10^{10}$} & {831466.91}        \\
\multicolumn{1}{c}{\multirow{4}*{Case III}} &  Frequency  & {0.00230322+}   &   {-1.00231+}   &   {-1.00101+}    &  {0.00241948+}     \\
\multicolumn{1}{c}{}                        &   (cpsd)    & {i0.0000134621} & {i0.0000183397} & {i0.00109607}   &  {i$2.63826\times10^{-6}$} \\
\multicolumn{1}{c}{}                        &  {Period}   &    {432.99}     &    {-431.72}     &     {-987.40}     &     {412.18}        \\
\multicolumn{1}{c}{}                        &     {Q}     &     {85.54}     &   {27326.24}    &       {456.64}     &      {458.54}        \\
\multicolumn{1}{c}{\multirow{4}*{Case IV}}  &  Frequency  & {0.00230323+}   &   {-1.00228+}   & \multirow{2}*{-1.0}    &  {0.00242191+}         \\
\multicolumn{1}{c}{}                        &   (cpsd)    & {i0.0000134552} & {i$3.11692\times10^{-8}$} & {}        &  {i$1.21192\times10^{-8}$}  \\
\multicolumn{1}{c}{}                        &  {Period}   &    {432.99}     &    {-437.40}    &         ——        &     {411.77}        \\
\multicolumn{1}{c}{}                        &     {Q}     &     {85.59}     & {16078051.41}   &         ——        &    {99920.37}     \\
\multicolumn{1}{c}{\multirow{4}*{Case V}}   
                                            &  Frequency  & {0.00230333+}   &  {-1.00229+}    &  {-0.99762+}    &  {0.000130128+}      \\
\multicolumn{1}{c}{}                        &   (cpsd)    &{i0.0000134673}  & {i$3.14366\times10^{-8}$} & {i$4.62954\times10^{-11}$}   & {i$1.17229\times10^{-13}$} \\
\multicolumn{1}{c}{}                        &  {Period}   &    {432.97}     &    {-435.49}    &    {419.02}     &     {7663.76}        \\
\multicolumn{1}{c}{}                        &     {Q}     &     {85.52}     &   {15994976.29} &     {$1.0775086\times10^{10}$}     &      {$5.5501625\times10^{8}$}        \\
\multicolumn{1}{c}{\multirow{4}*{Case VI}}  &  Frequency  & {0.00230158+}   &  {-1.00228+}     &  {-1.00524+}    &  {0.00765573+}      \\
\multicolumn{1}{c}{}                        &   (cpsd)    &{i0.0000134521}  & {i$3.11364\times10^{-8}$} & {i$1.94927\times10^{-10}$} & {i$1.51279\times10^{-8}$}    \\
\multicolumn{1}{c}{}                        &  {Period}   &    {433.30}     &    {-437.40}     &      {-190.32}    &     {130.26}        \\
\multicolumn{1}{c}{}                        &     {Q}     &     {85.55}     & {16149136.38}   & {$2.5785037\times10^{9}$} & {253033.47}        \\
\multicolumn{1}{c}{\multirow{4}*{Case VII}} &  Frequency  & {0.00230332+}   &   {-1.00232+}   &   {-0.998635+}    &  {0.000129996+}     \\
\multicolumn{1}{c}{}                        &   (cpsd)    & {i0.0000134725} & {i0.0000223634} & {i0.00109454}   &  {i$1.41861\times10^{-7}$} \\
\multicolumn{1}{c}{}                        &  {Period}   &    {432.97}     &    {-429.86}     &     {730.60}     &     {7671.54}        \\
\multicolumn{1}{c}{}                        &     {Q}     &     {85.48}     &   {22409.83}    &     {456.19}     &      {458.18}        \\
\multicolumn{1}{c}{\multirow{4}*{Case VIII}}&  Frequency  & {0.00230157+}   &   {-1.00232+}   & {-1.00623+}       &  {0.00764811+}         \\
\multicolumn{1}{c}{}                        &   (cpsd)    & {i0.0000134601} & {i0.0000285493}  & {i0.00108028}    &  {i$8.22752\times10^{-6}$}  \\
\multicolumn{1}{c}{}                        &  {Period}   &    {433.30}     &    {-429.86}     &     {-160.08}     &     {130.39}        \\
\multicolumn{1}{c}{}                        &     {Q}     &     {85.50}     &   {17554.20}     &     {465.73}      &     {464.79}     \\
\specialrule{0.1em}{0.25pt}{0.25pt}
\end{tabular}
\vspace{2mm}
\end{table}

The CW periods in Cases I and II were 432.92 d and 432.918 d, respectively, the lengthening of 0.002 d was caused by electromagnetic couplings near the CMB and ICB. Thus, we can conclude that electromagnetic couplings have a negligible effect on the CW and can also been found by comparing Cases III and IV, or Cases V and VII, or Cases VI and VIII. The CW periods in Cases III and IV were both 432.99 d, indicating that the pressure and gravitational coupling shorten the CW by about 0.07 d. In Cases V and VII, the CW periods were both 432.97 d, therefore, the absent gravitational coupling shortened the CW by about 0.05 d compared to Case II. In Cases VI and VIII, the CW periods were both 433.30 d, indicating that the absent pressure coupling shortened the CW by about 0.38 d compared to Case II. After considering the pressure coupling damping effect, namely multiplying the damping factor $\alpha_3$, the absent pressure coupling shortened the CW about 0.02 d. Thus, consistent with the above result, the total effect of pressure and gravitational coupling will shorten the CW by about 0.07 d.

The FCN periods in Cases I and II were $-429.86$ d and $-435.49$ d, respectively, the electromagnetic couplings therefore shortened the FCN by about 5.63 d when comparing Case I with Case II. The FCN periods in Cases II and V were both $-435.49$ d, and the FCN periods in Cases IV and VI were both $-435.49$ d. This indicates that gravitational coupling acting alone on the solid inner core has no effect on the FCN. By comparing Cases II and V with Case IV, we found that the pressure coupling can shorten the FCN by about 1.91 d. By comparing Case III with Case IV, the electromagnetic coupling shortened the FCN by about 5.68 d. When comparing Case V with Case VII, we found that the electromagnetic coupling will shorten the FCN by about 5.63 d. However, when comparing Case VI with Case VIII, the electromagnetic coupling shortened the FCN by about 7.54 d. The inconsistent result may be caused by different interactions between the pressure and viscoelectromagnetic couplings, respectively, with the gravitational and viscoelectromagnetic couplings acting on the solid inner core.

FICN and ICW are two rotational normal modes of the solid inner core, and their behaviors are different from each other. In case I, the FICN and ICW periods were 921.69 d and 2413.01 d, respectively, representing the comprehensive results of the pressure, gravitational, and viscoelectromagnetic couplings and the role of each coupling is obscured. Hence, it is our motivation to design the different cases above to investigate the individual roles of the pressure, gravitational, and viscoelectromagnetic couplings. 

We found that the pressure coupling dominates the FICN and causes its prograde motion, while the viscoelectromagnetic and gravitational couplings will lead to a retrograde FICN motion. \citet{Xu1998} used the Euler kinematic and dynamical equation to investigate the effect of pressure and gravitational coupling on the FICN and ICW, and found that the FICN can be retrograde when there is no pressure coupling. When modeling the electromagnetic couplings, \citet{buffett2002} also found that the electromagnetic couplings will cause the FICN to be retrograde. Similarly, in studying the forced precession of the inner core of the Moon, \citet{dumberry2016} stated that the pressure coupling dominates over the gravitational coupling and leads to a prograde FICN for the Earth. Therefore, the FICN period was 419.02 d for Case V, 475.57 d for Case II, 730.60 d for Case VII, and 921.69 d for Case I, where the gravitational and viscoelectromagnetic couplings partly cancel the dominant pressure coupling and lead to a longer FICN period. If there is no pressure coupling, under the control of viscoelectromagnetic couplings or gravitational coupling, the FICN will be retrograde, with a period of $-987.40$ d as in Case III, $-190.32$ d in Case VI, and $-160.08$ d in Case VIII. If there is no pressure coupling, the gravitational coupling and viscoelectromagnetic coupling acting on the solid inner core, then the FICN frequency will be $-1.0$ cpsd, indicating the absence of the FICN first found by \citet{Xu1998}. 

We also found that the pressure and gravitational couplings were more important for the ICW than the viscoelectromagnetic couplings, that only had a slight effect and can be seen by comparing Cases I and II, Cases III and IV, Cases V and VII, and Cases VI and VIII. The pressure coupling is a damping torque and the main cause for long ICW periods. Hence, the ICW period was about 130.0 d in Cases VI and VIII, 7663.76 d (21 yr) in Case V, and 7671.54 d (21 yr) in Case VII. If both pressure and gravitational coupling are considered, the gravitational coupling will also be damped by the fluid outer core with a damping factor of $\alpha_3$, and thus the ICW period was 2410.56 d or 2413.01 d (6.6 yr), where the viscoelectromagnetic couplings only slightly lengthened the ICW. If there is no pressure and gravitational coupling, the ICW period will be about 412.0 d, and represents the essence of the dynamical ellipticity of the solid inner core first pointed out by \citet{mathews1991a}. \citet{Xu1998} also investigated the effect of pressure and gravitational coupling on ICW in the time domain and found that the ICW period was 132 d only in the gravitational coupling case, 416 d in the case without coupling on the inner core, 5 yr in the case with both pressure and gravitational coupling, and 20 yr in the pressure only coupling case. These results are close to our results and the differences may be caused by the rigid mantle and inner core Earth model they deployed.

Further numerical calculations of the rotational normal modes indicated that the viscoelectromagnetic coupling near the CMB had hardly any effect on the FICN and ICW. The pressure, gravitational, and viscoelectromagnetic couplings acting on the solid inner core dominated the rotation behaviors of the FICN and ICW.

\subsection{Comparison with Other Results}
\label{S6.2}
\citet{escapa2001} constructed a rotationally symmetric three-layered Earth rotation theory and provided four rotational normal mode solutions based on a rigid mantle, fluid outer core, and rigid inner core Earth model by using the HA approach. The normal mode solutions of \citet{escapa2001} are very different from those of \citet{mathews1991b} because they used different Earth models, but consistent results could be obtained when simplifying the MBHS1991 theory \citep{mathews1991a}. Under the assumption that the Earth consists of a rigid mantle, fluid outer core, and rigid inner core, considering only the pressure coupling and the rotationally symmetric case, our theory provides normal mode solutions that are close to the results given by the rotationally symmetric three-layered Earth rotation theory constructed by \citet{escapa2001}. \citet{escapa2001} provided the following values for the above pressure coupling parameters:
\begin{equation}
\label{58}
\alpha_1=\alpha_2=\frac{\delta}{e_s},\quad \alpha_3=1-\frac{\delta}{e_s}
\end{equation}
where $\delta$ equals $0.002232$. When we deployed the dynamic figure parameters in \citet{escapa2001} for comparison purposes, we obtained rotationally normal mode solutions very close to their numerical results both with the trigometric function and eigenvalue methods, as listed in Table \ref{T6}.

\begin{table}[htp]
\centering
\caption{Comparison of the rotational normal mode solutions in this study with the results of \citet{escapa2001}}
\label{T6}
\begin{tabular}{ccccc}
\specialrule{0.1em}{0.5pt}{0.5pt}
                  &       CW       &       FCN      &       FICN      &         ICW        \\
\specialrule{0.05em}{0.25pt}{0.25pt}
Frequency (cpsd)  &  0.00366845    &   -1.00288     &    -0.997761    &   0.000173387      \\
Period (d)        &     271.85     &     -346.27    &      445.41     &     5751.70        \\
Q value           &       ——       &       ——       &       ——        &        ——          \\
\citet{escapa2001} &     271.85     &     -346.03    &      445.38     &     5751.70        \\
\specialrule{0.1em}{0.25pt}{0.25pt}
\end{tabular}
\vspace{2mm}
\end{table}

If we neglect all terms related to solid inner core, our theory is simplified to triaxial two-layered Earth rotation theory as formulated by \citet{chen2010}, that can be expressed as:
\begin{equation}
\begin{gathered}
\label{59}
(1+\kappa)A\frac{\mathrm{d}m_1}{\mathrm{d}t}+(A_f+A\xi)\frac{\mathrm{d}m_1^f}{\mathrm{d}t}+(C-(1+\kappa)B)\Omega_0m_2\\
-(B_f+B\xi)\Omega_0m_2^f = A\kappa\frac{\mathrm{d}\phi_1}{\mathrm{d}t}+(C-(1+\kappa)B)\Omega_0 \phi_2\\
(1+\kappa)B\frac{\mathrm{d}m_2}{dt}+(B_f+B\xi)\frac{\mathrm{d}m_2^f}{\mathrm{d}t}-(C-(1+\kappa)A)\Omega_0m_1\\
+(A_f+A\xi)\Omega_0m_1^f = B\kappa\frac{\mathrm{d}\phi_2}{\mathrm{d}t}-(C-(1+\kappa)A)\Omega_0 \phi_1\\
A_f(1+\gamma)\frac{\mathrm{d}m_1}{\mathrm{d}t}+A_f(1+\beta)\frac{\mathrm{d}m_1^f}{\mathrm{d}t}-\Omega_0(C_f+K_{CMB}B_f)m_2^f = A_f \gamma \frac{\mathrm{d}\phi_1}{\mathrm{d}t}\\
B_f(1+\gamma)\frac{\mathrm{d}m_2}{\mathrm{d}t}+B_f(1+\beta)\frac{\mathrm{d}m_2^f}{\mathrm{d}t}+\Omega_0(C_f+K_{CMB}A_f)m_1^f = B_f \gamma \frac{\mathrm{d}\phi_2}{\mathrm{d}t}
\end{gathered}
\end{equation}
where the variables corresponding to the former fluid outer core now represent the variables for the whole fluid core, namely the outer core and inner core constituting the whole fluid core. If we deploy the trigometric function method, the normal mode matrix $M$ now degenerates to a $8\times8$ matrix:
\begin{equation*}
\bm{M}=\left[
\begin{array}{cccc}
0&A(1+\kappa)\sigma&C-(1+\kappa)B&0 \\
-A(1+\kappa)\sigma&0&0&C-(1+\kappa)B \\
-(C-(1+\kappa)A)&0&0&B(1+\kappa)\sigma \\
0&-(C-(1+\kappa)A)&-B(1+\kappa)\sigma&0 \\
0&A_f(1+\gamma)\sigma&0&0 \\
-A_f(1+\gamma)\sigma&0&0&0 \\
0&0&0&B_f(1+\gamma)\sigma \\
0&0&-B_f(1+\gamma)\sigma&0
\end{array}\right.
\end{equation*}
\begin{equation}
\label{60}
\left.\begin{array}{cccc}
0&(A_f+A\xi)\sigma&-(B_f+B\xi)&0 \\
-(A_f+A\xi)\sigma&0&0&-(B_f+B\xi) \\
A_f+A\xi&0&0&(B_f+B\xi)\sigma \\
0&A_f+A\xi&-(B_f+B\xi)\sigma&0 \\
0&A_f(1+\beta)\sigma&-(C_f+B_fK_{CMB})&0 \\
-A_f(1+\beta)\sigma&0&0&-(C_f+B_fK_{CMB}) \\
C_f+A_fK_{CMB}&0&0&B_f(1+\beta)\sigma \\
0&C_f+A_fK_{CMB}&-B_f(1+\beta)\sigma&0
\end{array}
\right]
\end{equation}
and the normal modes of the anelastic mantle, fluid core Earth model, namely the CW and FCN, can be derived by setting $|M|=0$. If we deploy the eigenvalue method, the triaxial two-layered Earth rotation theory can be written into matrix form as:
\begin{equation}
\label{61}
\mathbf{F}\frac{\mathrm{d}\mathbf{y}}{\mathrm{d}t}=\Omega_0 \mathbf{G}\mathbf{y},
\end{equation}
where
\begin{equation}
\label{62}
\mathbf{y}=\big[m_1\ m_2\ m^f_1\ m^f_2\big]^T,
\end{equation}
and
\begin{equation}
\label{63}
\mathbf{F}=\left[
\begin{array}{cccc}
A(1+\kappa)&0&A_f+A\xi&0 \\
0&B(1+\kappa)&0&B_f+B\xi \\
A_f(1+\gamma)&0&A_f(1+\beta)&0 \\
0&B_f(1+\gamma)&0&B_f(1+\beta)
\end{array}\right],
\end{equation}
and
\begin{equation}
\label{64}
\mathbf{G}=\left[
\begin{array}{cccc}
0&-(C-(1+\kappa)B)&0&B_f+B\xi \\
C-(1+\kappa)A&0&-(A_f+A\xi)&0 \\
0&0&0&C_f+B_f K_{CMB} \\
0&0&-(C_f+A_f K_{CMB})&0
\end{array}\right].
\end{equation}
Similarly, the CW and FCN rotational normal modes of this triaxial two-layered Earth rotation theory can be obtained by solving the eigenvalue $\sigma_i=\mathbf{F^{-1}}\mathbf{G}$ and the rotational normal mode solutions will be $-i\sigma_i$ as stated in section \ref{S3.2}.

Here, we used the data in Table \ref{T1} with the trigometric function and eigenvalue methods to calculate the rotational normal modes of this triaxial two-layered Earth rotation theory, namely CW and FCN, the results of which were the same, and consistent with those listed in Table \ref{T7}. The period and quality factor of the CW were very close to those as given in Case I (see section \ref{S5}), and the period and quality factor of the FCN were larger than those given in Case I (see section \ref{S5}) due to the fact that this two-layered case did not consider the viscoelectromagnetic coupling, and the pressure and gravitational couplings acting on the solid inner core. 

\begin{table}[htp]
\centering
\caption{Rotational normal modes for the triaxial two-layered Earth model}
\label{T7}
\begin{tabular}{ccc}
\specialrule{0.1em}{0.5pt}{0.5pt}
                  &       CW       &       FCN              \\
\specialrule{0.05em}{0.25pt}{0.25pt}
Frequency         &  0.00230199+   &   -1.00231+      \\
(cpsd)            & i0.0000134729  & i0.0000201055    \\
Period (d)        &     433.22     &     -431.72      \\
Q value           &      85.43     &    24926.26       \\
\citet{chen2010}   &  433.03/100.20 &   430.34/——      \\
\specialrule{0.1em}{0.25pt}{0.25pt}
\end{tabular}
\end{table}

\section{Conclusions}
 As a generalization of both the rotationally symmetric three-layered Earth rotation theory (MBHS1991 theory) and the triaxial two-layered Earth rotation theory \citep{chen2010}, here we formulated a triaxial three-layered Earth rotation theory. Our study showed that the triaxiality will lead to a CW period increase of about 0.01 d, and an ICW period increase of about 0.35 d.

There are three kinds of dissipative processes; mantle anelasticity induced dissipation, ocean tide induced dissipation, and viscoelectromagnetic coupling induced dissipation. The mantle anelasticity and ocean tide induced dissipations are mainly responsible for the CW, and contribute little to the FCN, while the viscoelectromagnetic coupling induced dissipations are mainly responsible for the FCN, FICN, and ICW. 

By fitting the calculated results to the CW and FCN observations, a group of new compliances were obtained, providing $\kappa=1.24063\times10^{-3}-1.195\times10^{-5}i$, $\gamma=1.601\times10^{-3}-1.073\times10^{-4}i$, and $\beta=6.238\times10^{-4}$, with other the compliances unchanged. Using these new compliances, new normal mode solutions can be provided under the rotationally symmetric case, allowing us to validate the results from the triaxial case with those from the biaxial case. Compared to the biaxial case, in the triaxial case, the CW period increased by about 0.01 d, the ICW period was prolonged by about 0.35 d, and the triaxiality had no effect on the FCN and FICN in the current parameter settings. 

Conventionally, the viscoelectromagnetic couplings, pressure and gravitational coupling are combined when investigating their effects on the rotational normal modes, and this may conceal the individual role of core mantle couplings. Hence, we separated the pressure coupling from the gravitational coupling, and provided various numerical results for the rotational normal modes using eight different cases. 

The viscoelectromagnetic couplings had a nearly negligible effect on the CW, while the pressure and gravitational coupling shortened the CW by about 0.07 d. The gravitational coupling either individually, or combined with the pressure coupling had no effect on the FCN, however, the pressure coupling shortened the FCN by about 1.91 d. The viscoelectromagnetic couplings either individually, combined with the pressure coupling, or combined with the pressure and gravitational coupling shortened the FCN by about 5.6 d, while combining the viscoelectromagnetic couplings with the gravitational coupling shortened the FCN by about 7.5 d. This phenomenon might be caused by different interactions between the viscoelectromagnetic coupling with the pressure or gravitational couplings acting on the solid inner core, respectively, and may be evidence for the effect of the solid inner core on the FCN.

Pressure coupling dominates a prograde FICN, while the viscoelectromagnetic and gravitational couplings will lead to a retrograde FICN and partly cancel the effect of the pressure coupling to give rise to a longer FICN period. The viscoelectromagnetic couplings will lengthen the ICW slightly, while the pressure coupling will damp and cause a longer ICW period. Considering only the gravitational coupling, the ICW period will be about 130 d, which is very short, while considering the damping of the pressure coupling, the ICW period will be 6.6 yrs. 

 Our study shows that, under different conditions, the triaxial three-layered Earth rotation theory will degenerate to the rotationally symmetric three-layered Earth rotation theory or the triaxial two-layered Earth rotation theory. Here we point out that, in this study, we did not consider the second-order effects of geometric and dynamical flattenings of the triaxial three-layered Earth model, or the topographic couplings between the inner and outer cores and between the mantle and outer core, and these will be investigated in future studies.

\justifying

\acknowledgments
We would like to express our sincere thanks to W. Chen and H. S. Fok for discussions that improved this manuscript. This study was supported by the NSFC (grant Nos. 41631072, 41721003, 41874023, 41574007, and 41429401), the Discipline Innovative Engineering Plan of Modern Geodesy and Geodynamics (grant No. B17033), and the DAAD Thematic Network Project (grant No. 57173947)

%% ------------------------------------------------------------------------ %%
%% References and Citations

%%%%%%%%%%%%%%%%%%%%%%%%%%%%%%%%%%%%%%%%%%%%%%%
% BibTeX is preferred:
%
% \bibliography{<name of your .bib file>}
%
% don't specify bibliographystyle
%%%%%%%%%%%%%%%%%%%%%%%%%%%%%%%%%%%%%%%%%%%%%%%

% Please use ONLY \citet and \citep for reference citations.
% DO NOT use other cite commands (e.g., \cite, \citeyear, \nocite, \citealp, etc.).
%% Example \citet and \citep:
%  ...as shown by \citet{Boug10}, \citet{Buiz07}, \citet{Fra10},
%  \citet{Ghel00}, and \citet{LeiT64}.

%  ...as shown by \citep{Boug10}, \citep{Buiz07}, \citep{Fra10},
%  \citep{Ghel00, LeiT64}.

%  ...has been shown \citep [e.g.,][]{Boug10,Buiz07,Fra10}.

\bibliography{zhlgref}

\appendix
\section{The Normal Mode Matrix using the Trigometric Function Method}}
\label{A}
Here, we provide the normal mode matrix using the trigometric function method in different core mantle coupling conditions that can be categorized into four classes:\\
\textbf{Class I} Pressure and gravitational couplings plus viscoelectromagnetic couplings;\\
\textbf{Class II} Pressure plus viscoelectromagnetic couplings;\\
\textbf{Class III} Gravitational plus viscoelectromagnetic couplings;\\
\textbf{Class IV} Without pressure and gravitational couplings plus Only viscoelectromagnetic couplings.\\
The viscoelectromagnetic couplings can be easily set to zero to generate the other core mantle coupling cases in section \ref{S6.1}.

For \textbf{Class I}, the element $4\times 4$ matrix $M_{ij}\,(i,j=1,2,3,4)$ of the normal mode $16\times16$ matrix $\bm{M}$ of our triaxial three-layered Earth rotation theory can be expressed as: 
\begin{equation}
\label{A.1}
\begin{aligned}
M_{11}=\left[
\begin{array}{cccc}
0&(1+\kappa)\sigma A&C-(1+\kappa)B&0\\
-(1+\kappa)\sigma A&0&0&C-(1+\kappa)B\\
-(C-(1+\kappa)A)&0&0&(1+\kappa)\sigma B\\
0&-(C-(1+\kappa)A)&-(1+\kappa)\sigma B&0
\end{array}
\right]
\end{aligned}
\end{equation}

\begin{equation}
\label{A.2}
\begin{aligned}
M_{12}=\left[
\begin{array}{cccc}
0&(A_f+A\xi)\sigma&-(B_f+B\xi)&0\\
-(A_f+A\xi)\sigma&0&0&-(B_f+B\xi)\\
A_f+A\xi&0&0&(B_f+B\xi)\sigma\\
0&A_f+A\xi&-(B_f+B\xi)\sigma&0
\end{array}
\right]
\end{aligned}
\end{equation}

\begin{equation}
\label{A.3}
\begin{aligned}
M_{13}=\left[
\begin{array}{cccc}
0&(A_s+A\varsigma)\sigma&-(B_s+B\varsigma)&0\\
-(A_s+A\varsigma)\sigma&0&0&-(B_s+B\varsigma)\\
A_s+A\varsigma&0&0&(B_s+B\varsigma)\sigma\\
0&A_s+A\varsigma&-(B_s+B\varsigma)\sigma&0
\end{array}
\right]
\end{aligned}
\end{equation}

\begin{equation}
\label{A.4}
\begin{aligned}
M_{14}=\left[
\begin{array}{cccc}
0&(A_s+B_s)\alpha_3 e_s\sigma/2&-(A_s+B_s)\alpha_3 e_s/2&0\\
-(A_s+B_s)\alpha_3 e_s \sigma/2&0&0&-(A_s+B_s)\alpha_3 e_s/2\\
(A_s+B_s)\alpha_3 e_s/2&0&0&(A_s+B_s)\alpha_3 e_s \sigma/2\\
0&(A_s+B_s)\alpha_3 e_s/2&-(A_s+B_s)\alpha_3 e_s \sigma/2&0
\end{array}
\right]
\end{aligned}
\end{equation}

\begin{equation}
\label{A.5}
\begin{aligned}
M_{21}=\left[
\begin{array}{cccc}
0&A_f(1+\gamma)\sigma&0&0\\
-A_f(1+\gamma)\sigma&0&0&0\\
0&0&0&B_f(1+\gamma)\sigma\\
0&0&-B_f(1+\gamma)\sigma&0
\end{array}
\right]
\end{aligned}
\end{equation}

\begin{equation}
\label{A.6}
\begin{aligned}
M_{22}=\left[
\begin{array}{cc}
0&A_f(1+\beta)\sigma\\
-A_f(1+\beta)\sigma&0\\
C_f+K_{CMB}A_f+K_{ICB}A_s&0\\
0&C_f+K_{CMB}A_f+K_{ICB}A_s
\end{array}\right.\\
\left.\begin{array}{cc}
-(C_f+K_{CMB}B_f+K_{ICB}B_s)&0 \\
0&-(C_f+K_{CMB}B_f+K_{ICB}B_s) \\
0&B_f(1+\beta)\sigma \\
-B_f(1+\beta)\sigma&0
\end{array}
\right]
\end{aligned}
\end{equation}

\begin{equation}
\label{A.7}
\begin{aligned}
M_{23}=\left[
\begin{array}{cccc}
0&A_f\delta\sigma&K_{ICB}B_s&0\\
-A_f\delta\sigma&0&0&K_{ICB}B_s\\
-K_{ICB}A_s&0&0&B_f\delta\sigma\\
0&-K_{ICB}A_s&-B_f\delta\sigma&0
\end{array}
\right]
\end{aligned}
\end{equation}

\begin{equation}
\label{A.8}
\begin{aligned}
M_{24}=\left[
\begin{array}{cccc}
0&-(A_s+B_s)\alpha_1 e_s\sigma/2&0&0\\
(A_s+B_s)\alpha_1 e_s\sigma/2&0&0&0\\
0&0&0&-(A_s+B_s)\alpha_1 e_s\sigma/2\\
0&0&(A_s+B_s)\alpha_1 e_s\sigma/2&0
\end{array}
\right]
\end{aligned}
\end{equation}

\begin{equation}
\label{A.9}
\begin{aligned}
M_{31}=\left[
\begin{array}{cccc}
0&A_s(1+\theta)\sigma&(C_s-B_s)(1-\alpha_1)&0\\
-A_s(1+\theta)\sigma&0&0&(C_s-B_s)(1-\alpha_1)\\
-(C_s-A_s)(1-\alpha_1)&0&0&B_s(1+\theta)\sigma\\
0&-(C_s-A_s)(1-\alpha_1)&-B_s(1+\theta)\sigma&0
\end{array}
\right]
\end{aligned}
\end{equation}

\begin{equation}
\label{A.10}
\begin{aligned}
M_{32}=\left[
\begin{array}{cc}
0&\chi A_s\sigma\\
-\chi A_s\sigma&0\\
(C_s-A_s)\alpha_1-A_s K_{ICB}&0\\
0&(C_s-A_s)\alpha_1-A_s K_{ICB}
\end{array}
\right.\\
\left.\begin{array}{cc}
-[(C_s-B_s)\alpha_1-B_s K_{ICB}]&0\\
0&-[(C_s-B_s)\alpha_1-B_s K_{ICB}]\\
0&\chi B_s\sigma\\
-\chi B_s\sigma&0
\end{array}
\right]
\end{aligned}
\end{equation}

\begin{equation}
\label{A.11}
\begin{aligned}
M_{33}=\left[
\begin{array}{cccc}
0&A_s(1+\nu)\sigma&-B_s(1+K_{ICB})&0\\
-A_s(1+\nu)\sigma&0&0&-B_s(1+K_{ICB})\\
A_s(1+K_{ICB})&0&0&B_s(1+\nu)\sigma\\
0&A_s(1+K_{ICB})&-B_s(1+\nu)\sigma&0
\end{array}
\right]
\end{aligned}
\end{equation}

\begin{equation}
\label{A.12}
\begin{aligned}
M_{34}=\left[
\begin{array}{cc}
0&(A_s+B_s)e_s\sigma/2\\
-(A_s+B_s)e_s\sigma/2&0\\
(A_s+B_s)e_s/2-\alpha_2(C_s-A_s)&0\\
0&(A_s+B_s)e_s/2-\alpha_2(C_s-A_s)
\end{array}\right.\\
\left.\begin{array}{cc}
-((A_s+B_s)e_s/2-\alpha_2(C_s-B_s))&0\\
0&-((A_s+B_s)e_s/2-\alpha_2(C_s-B_s))\\
0&(A_s+B_s)e_s\sigma/2\\
-(A_s+B_s)e_s\sigma/2&0
\end{array}
\right]
\end{aligned}
\end{equation}

\begin{equation}
\label{A.13}
\begin{aligned}
M_{41}=\left[
\begin{array}{cccc}
0&0&0&0\\
0&0&0&0\\
0&0&0&0\\
0&0&0&0
\end{array}\right],\ 
M_{42}=
\left[\begin{array}{cccc}
0&0&0&0\\
0&0&0&0\\
0&0&0&0\\
0&0&0&0
\end{array}
\right]
\end{aligned}
\end{equation}

\begin{equation}
\label{A.14}
\begin{aligned}
M_{43}=\left[
\begin{array}{cccc}
0&0&-1&0\\
0&0&0&-1\\
1&0&0&0\\
0&1&0&0
\end{array}\right],\ 
M_{44}=\left[
\begin{array}{cccc}
0&\sigma&0&0\\
-\sigma&0&0&0\\
0&0&0&\sigma\\
0&0&-\sigma&0
\end{array}
\right].
\end{aligned}
\end{equation}

Note that for \textbf{Class II, III, IV}, only the components related with solid inner core are different from components of \textbf{Class I}, hence, we provided only the components of the normal mode matrix related with the solid inner core below.

For \textbf{Class II}, the first three components of the normal mode matrix related with the solid inner core, i.e. $M_{3j}\,(j=1,2,3)$ are the same as \textbf{Class I}, and the only different component $M_{34}$ can be expressed as:
\begin{equation}
\label{A.15}
\begin{aligned}
M_{34}=\left[
\begin{array}{cc}
0&(A_s+B_s)e_s\sigma/2\\
-(A_s+B_s)e_s\sigma/2&0\\
(A_s+B_s)e_s/2-\alpha_1(C_s-A_s)&0\\
0&(A_s+B_s)e_s/2-\alpha_1(C_s-A_s)
\end{array}\right.\\
\left.\begin{array}{cc}
-((A_s+B_s)e_s/2-\alpha_1(C_s-B_s))&0\\
0&-((A_s+B_s)e_s/2-\alpha_1(C_s-B_s))\\
0&(A_s+B_s)e_s\sigma/2\\
-(A_s+B_s)e_s\sigma/2&0
\end{array}
\right].
\end{aligned}
\end{equation}
We found that the \textbf{Class II} can be obtained by setting $\alpha_g=0,\ \alpha_2=\alpha_1-\alpha_3 \alpha_g=\alpha_1$ in \textbf{Class I}.

For \textbf{Class III}, the component of the normal mode matrix related with the solid inner core (i.e. $M_{3j},\ j=1,2,3,4$) can be expressed as:
\begin{equation}
\label{A.16}
\begin{aligned}
M_{31}=\left[
\begin{array}{cccc}
0&A_s(1+\theta)\sigma&C_s-B_s(1+\theta)&0\\
-A_s(1+\theta)\sigma&0&0&C_s-B_s(1+\theta)\\
-(C_s-A_s(1+\theta))&0&0&B_s(1+\theta)\sigma\\
0&-(C_s-A_s(1+\theta))&-B_s(1+\theta)\sigma&0
\end{array}
\right]
\end{aligned}
\end{equation}

\begin{equation}
\label{A.17}
\begin{aligned}
M_{32}=\left[
\begin{array}{cccc}
0&\chi A_s\sigma&B_s(K_{ICB}-\chi)&0\\
-\chi A_s\sigma&0&0&B_s(K_{ICB}-\chi)\\
-A_s(K_{ICB}-\chi)&0&0&\chi B_s\sigma\\
0&-A_s(K_{ICB}-\chi)&-\chi B_s\sigma&0
\end{array}
\right]
\end{aligned}
\end{equation}

\begin{equation}
\label{A.18}
\begin{aligned}
M_{33}=\left[
\begin{array}{cccc}
0&A_s(1+\nu)\sigma&-B_s(1+K_{ICB}+\nu)&0\\
-A_s(1+\nu)\sigma&0&0&-B_s(1+K_{ICB}+\nu)\\
A_s(1+K_{ICB}+\nu)&0&0&B_s(1+\nu)\sigma\\
0&A_s(1+K_{ICB}+\nu)&-B_s(1+\nu)\sigma&0
\end{array}
\right]
\end{aligned}
\end{equation}

\begin{equation}
\label{A.19}
\begin{aligned}
M_{34}=\left[
\begin{array}{cc}
0&(A_s+B_s)e_s\sigma/2\\
-(A_s+B_s)e_s\sigma/2&0\\
(A_s+B_s)e_s/2+\alpha_g(C_s-A_s)&0\\
0&(A_s+B_s)e_s/2+\alpha_g(C_s-A_s)
\end{array}\right.\\
\left.\begin{array}{cc}
-((A_s+B_s)e_s/2+\alpha_g(C_s-B_s))&0\\
0&-((A_s+B_s)e_s/2+\alpha_g(C_s-B_s))\\
0&(A_s+B_s)e_s\sigma/2\\
-(A_s+B_s)e_s\sigma/2&0
\end{array}
\right].
\end{aligned}
\end{equation}

For \textbf{Class IV}, the first three components of the normal mode matrix related with the solid inner core, i.e. $M_{3j}\,(j=1,2,3)$ are the same as \textbf{Class III}, and the only different component $M_{34}$ can be expressed as:
\begin{equation}
\label{A.20}
\begin{aligned}
M_{34}=\left[
\begin{array}{cccc}
0&(A_s+B_s)e_s\sigma/2&-(A_s+B_s)e_s/2&0\\
-(A_s+B_s)e_s\sigma/2&0&0&-(A_s+B_s)e_s/2\\
(A_s+B_s)e_s/2&0&0&(A_s+B_s)e_s\sigma/2\\
0&(A_s+B_s)e_s/2&-(A_s+B_s)e_s\sigma/2&0
\end{array}
\right].
\end{aligned}
\end{equation}
We can find that the \textbf{Class IV} can be obtained by setting $\alpha_g=0$ in \textbf{Class III}.

\section{The Eigenvalue Matrix with Eigenvalue Method}
\label{B}
The core mantle coupling category cases in Appendix A are also correct for the normal mode solutions of the triaxial three-layered Earth rotation theory based on the eigenvalue method. The matrix $\mathbf{F}$ is the same for both four classes of core mantle couplings, and the matrix $\mathbf{G}$ for \textbf{Class I} has been provided in section \ref{S3.2}. For the \textbf{Class II}, the matrix $\mathbf{G}$ will be the same as \textbf{Class I} after setting $\alpha_g=0,\ \alpha_2=\alpha_1-\alpha_3 \alpha_g=\alpha_1$.

For the \textbf{Class III}, the matrix $\mathbf{G}$ can be expressed as:
\begin{equation}
\label{B.1}
\begin{aligned}
\mathbf{G}=\left[
\begin{array}{cccc}
0&-(C-(1+\kappa)B)&0&B_f+B\xi \\
C-(1+\kappa)A&0&-(A_f+A\xi)&0 \\
0&0&0&C_f+K_{CMB}B_f+K_{ICB}B_s \\
0&0&-(C_f+K_{CMB}A_f+K_{ICB}A_s)&0 \\
0&-(C_s-B_s(1+\theta))&0&-B_s(K_{ICB}-\chi) \\
C_s-A_s(1+\theta)&0&A_s(K_{ICB}-\chi)&0 \\
0&0&0&0 \\
0&0&0&0 \\
\end{array}\right.\\
\left.\begin{array}{cccc}
0&B_s+B\varsigma&0&(A_s+B_s)\alpha_3 e_s/2 \\
-(A_s+A\varsigma)&0&-(A_s+B_s)\alpha_3 e_s/2&0 \\
0&-K_{ICB}B_s&0&0 \\
K_{ICB}A_s&0&0&0 \\
0&B_s(1+K_{ICB}+\nu)&0&(A_s+B_s)e_s/2+\alpha_g(C_s-B_s) \\
-A_s(1+K_{ICB}+\nu)&0&-[(A_s+B_s)e_s/2+\alpha_g(C_s-A_s)]&0 \\
0&1&0&0 \\
-1&0&0&0
\end{array}
\right].
\end{aligned}
\end{equation}

For the \textbf{Class IV}, the matrix $\mathbf{G}$ can be obtained by setting $\alpha_g=0$ in \textbf{Class III}.
\end{document}